\documentclass[pdflatex,sn-mathphys-num]{sn-jnl}
\usepackage{graphicx} % Required for inserting images

\usepackage{graphicx}%
\usepackage{multirow}%
\usepackage{amsmath,amssymb,amsfonts}%
\usepackage{amsthm}%
\usepackage{mathrsfs}%
\usepackage[title]{appendix}%
\usepackage{xcolor}%
\usepackage{textcomp}%
\usepackage{manyfoot}%
\usepackage{booktabs}%
\usepackage{algorithm}%
\usepackage{algorithmicx}%
\usepackage{algpseudocode}%
\usepackage{listings}%
\usepackage{hyperref}

\usepackage[most]{tcolorbox}
\tcbuselibrary{listings}
\definecolor{forkorange}{RGB}{230,120,40}

\newtcbinputlisting{\outputfile}[3]{%
  listing file={#3},
  listing only,
  title={#2},
  colframe=#1, colback=#1!4,
  coltitle=white, fonttitle=\bfseries\small,
  boxrule=0.6pt, arc=1.5mm,
  left=1.5mm, right=1.5mm, top=0.5mm, bottom=0.5mm,
  breakable,
  listing options={
    basicstyle=\ttfamily\scriptsize,
    columns=fullflexible, keepspaces=true,
    breaklines=true,
    breakindent=0pt, 
    breakautoindent=false,
    postbreak={}, 
    literate={—}{{--}}1 {’}{{'}}1 {“}{{``}}1 {”}{{''}}1  
  }
}

\theoremstyle{thmstyleone}%
\theoremstyle{thmstyletwo}%

\theoremstyle{thmstylethree}%

\begin{document}

\title[Article Title]{Human and Machine Annotations Predict Brain Responses to Naturalistic Movies Across Development}
% 
% Vision-Language Encoding of Social and Affective Brain Activity During MovieViewing


\author[1]{\fnm{Yiran} \sur{Huang}}\email{yh87@njit.edu}

\author[1]{\fnm{Zeyong} \sur{Zhang}}\email{zz63@njit.edu}

\author[2]{\fnm{Rounak} \sur{Rahman}}\email{rr744@njit.edu}

\author[3]{\fnm{Bharat} \sur{Biswal}}\email{bharat.biswal@njit.edu}

\author*[3]{\fnm{Xin} \sur{Di}}\email{xin.di@njit.edu}

\author*[1]{\fnm{Mengjia} \sur{Xu}}\email{mx6@njit.edu}

\affil*[1]{\orgdiv{Department of Data Science}, \orgname{New Jersey Institute of Technology}, \orgaddress{\street{University Heights}, \city{Newark}, \postcode{07102}, \state{NJ}, \country{USA}}}

\affil[2]{\orgdiv{Department of Biology}, \orgname{New Jersey Institute of Technology}, \orgaddress{\street{University Heights}, \city{Newark}, \postcode{07102}, \state{NJ}, \country{USA}}}

\affil[3]{\orgdiv{Department of Biomedical Engineering}, \orgname{New Jersey Institute of Technology}, \orgaddress{\street{University Heights}, \city{Newark}, \postcode{07102}, \state{NJ}, \country{USA}}}

% \affil[3]{\orgdiv{Department}, \orgname{Organization}, \orgaddress{\street{Street}, \city{City}, \postcode{610101}, \state{State}, \country{Country}}}


\abstract{
% Objective:

% mention fmri/age 
% age effect - mention trend - how old decrease
% point out which range significant 

% background

% LLM can understand human rating and predict brain signals
% LLM video captioning and human ratings predict brain response during movie watching
Naturalistic movie watching provides a powerful framework for studying dynamic brain function, but linking movie content to neural responses typically relies on labor-intensive human annotation. We investigated whether multimodal large language models (MLLMs) can provide scalable socio-affective and semantic annotations for naturalistic fMRI. We compared MLLM-generated theory-of-mind (ToM) and valence ratings with human ratings and with regional BOLD responses across three animated movies. For two of these movies, we combined machine-derived visual features, natural-language scene descriptions, and socio-affective ratings in an ROI-wise encoding model of fMRI data from 66 participants aged 5--22 years. MLLM ratings partially tracked human judgments, but their neural correspondence differed by construct and model: ToM showed the most consistent human--machine overlap in visual cortex, whereas valence correspondence was largely movie dependent. In the encoding model, visual features yielded the strongest and most widespread prediction, whereas scene descriptions were more spatially selective, concentrated in dorsal-attention and visual regions and extending into right lateral temporal cortex. Age-group comparisons suggested feature-specific developmental differences: scene-description encoding in dorsal-attention cortex was higher in the oldest group, whereas visual-feature encoding in salience/ventral-attention, limbic, and control networks peaked in the middle group. Together, these findings indicate that MLLM-generated annotations capture meaningful aspects of human judgments and neural responses, supporting MLLMs as a scalable complement to, rather than a replacement for, human annotation in naturalistic neuroimaging.

}




\maketitle
\section{Introduction}

Understanding how the human brain extracts meaning from complex, dynamic environments is a central goal of cognitive neuroscience. Naturalistic stimuli such as movies provide a powerful paradigm for studying this process, as they simultaneously engage perceptual, linguistic, and social-cognitive systems over extended timescales~\cite{hasson2004intersubject, nastase2021narratives}. Recent work further suggests that temporal context plays an important role in capturing neural responses during narrative video processing, particularly in higher-order integrative regions~\cite{jindal2026does}. Movie viewing also elicits reproducible, stimulus-specific patterns of dynamic functional connectivity across distributed brain regions, suggesting that different movie contents engage distinct patterns of large-scale neural interaction~\cite{di2022dynamic}. Temporal graph learning methods have also been developed to model dynamic functional connectivity~\cite{huang2025brainatcl}. Recent advances in voxel-wise and region-of-interest (ROI) encoding models have made it possible to relate time-resolved stimulus features to regional brain activity~\cite{huth2016natural}, offering a quantitative framework for probing how different levels of representation, from low-level visual properties to high-level semantic and social constructs, are mapped onto neural responses.

Among high-level cognitive constructs, Theory of Mind (ToM) and emotional valence are particularly relevant for naturalistic perception. ToM reflects the ability to infer others' mental states and has traditionally been associated with regions of the default mode network (DMN), including medial prefrontal and temporoparietal cortices~\cite{saxe2013people,frith2006neural,richardson2018development}. Valence, in contrast, captures the affective tone of a stimulus along a continuum from pleasant to unpleasant~\cite{russell1980circumplex}, and is typically linked to distributed limbic and affective systems~\cite{lindquist2012brain}. Both constructs are conventionally characterized through \emph{human subjective ratings}, in which observers report the mental states or affective tone they perceive in a stimulus~\cite{cowen2017self, bradley1994measuring, richardson2018development, jacoby2016localizing, skerry2015neural}. Such ratings, however, are costly to collect at scale, and are difficult or impossible to obtain from certain populations, such as young children, who often lack the verbal and metacognitive capacity to reliably self-report social-affective judgments~\cite{richardson2018development, wellman2001meta}, which limits their use precisely in the developmental contexts where they would be most informative.

Recent large language models (LLMs) increasingly produce judgments consistent with human mentalistic and social inference~\cite{strachan2024testing}. Building on this capacity, LLMs have been adopted as scalable annotators that often match or exceed trained human coders on labeling tasks~\cite{gilardi2023chatgpt}, and their judgments align closely with human ratings of affective content such as emotional valence~\cite{bojic2025comparing, alrasheed2025evaluating}, while recent work showed that GPT-4V ratings of social features~\cite{santavirta2025gpt} closely match human judgments and their neural correlates. Unlike these studies of brief clips and images, we examined moment-to-moment ToM and valence ratings from open-source MLLMs during continuous naturalistic movie viewing across development.
Together with multimodal vision-language models that can interpret naturalistic scenes directly from video, these advances suggest that MLLMs may provide a scalable and reproducible complement to human annotation, particularly in populations such as young children from whom reliable self-report can be difficult to obtain.
However, benchmarks such as CogGym reveal persistent gaps between model and human judgments~\cite{ying2026coggym}. More specifically, it remains less clear whether such human--machine correspondence extends to continuous socio-affective judgments during naturalistic movie viewing, particularly for constructs such as ToM and valence, and whether the corresponding neural patterns are consistent across movies and MLLMs.

% Building on this, we systematically compare human behavioral ratings with ratings generated by two state-of-the-art multimodal large language models (MLLMs), LLaDA-V~\cite{you2026llada} and Qwen3-VL~\cite{bai2025qwen3}, in their ability to explain ROI-level fMRI responses during naturalistic movie watching. We focus on ToM and valence across three movie stimuli (\emph{The Present}, \emph{Despicable Me}, and \emph{Partly Cloudy}), enabling assessment of both within-movie and cross-movie consistency. We investigate whether MLLM-derived social representations align with both human judgments and their underlying neural correlates.

% We further asked whether MLLMs representations contain information relevant to neural computation. This naturally motivates using MLLM-derived features within encoding models of brain activity. Our work then moves from behavior to prediction: can MLLM-derived features actively improve models of brain activity? 

A key challenge in encoding models lies in defining feature spaces that capture the richness of real-world experience. Traditional encoding approaches have used stimulus-derived representations spanning multiple levels, from low-level visual descriptors such as motion-energy and Gabor-wavelet features~\cite{kay2008identifying,nishimoto2011reconstructing}, to visual features extracted from pretrained convolutional neural networks~\cite{sohn2024explorations}, and further to higher-level semantic and social-cognitive features such as semantic categories, mentalization, and social interaction~\cite{huth2012continuous,lee2024multidimensional}. Although these approaches can capture both perceptual and higher-order social properties of naturalistic stimuli, they often depend on predefined feature categories and labor-intensive human annotation, limiting their scalability and the richness of the representations that can be examined.
Recent multimodal AI models have integrated visual, auditory, and linguistic information to predict neural responses across diverse settings~\cite{d2026foundation}, including parcel-wise fMRI responses during movie viewing~\cite{schad2025vibe}. Representations extracted from multimodal transformers yield encoding models that transfer across fMRI responses to language and vision~\cite{tang2023brain} and capture activity in social brain regions during natural movie viewing~\cite{small2024vision}. Related movie-fMRI encoding work has jointly modeled visual and language representations, showing that visual model features predict widespread cortical responses and contribute strongly even within individually localized social-interaction and language-selective regions~\cite{small2025ubiquitous}. AI-generated semantic descriptions have also been used to model ROI-level responses during movie viewing~\cite{li2026mapping}. Recent large-scale benchmarking further suggests that model representations that align with human social judgments do not necessarily align with human neural responses during dynamic social perception~\cite{garcia2025modeling}. 
% However, subject-specific encoding of machine-derived representations, particularly alongside direct comparisons between machine- and human-generated social annotations, remains less explored.
% Recent large-scale benchmarking further suggests that model representations that align with human social judgments do not necessarily align with human neural responses during dynamic social perception~\cite{garcia2025modeling}. 
However, subject-specific encoding of explicit MLLM-generated socio-affective and semantic representations remains less explored, particularly alongside direct comparisons between machine- and human-generated social annotations. Existing studies have also primarily evaluated latent representations extracted from pretrained models, leaving open whether explicit socio-affective judgments and semantic descriptions generated by modern MLLMs can capture both human judgments and brain activity.

% Building on this direction, our encoding framework extends that of~\cite{small2025ubiquitous} by incorporating MLLM-generated ratings and semantic descriptions to examine whether machine-derived features can account for ROI-level brain activity during movie watching and how they compare with human-derived ratings. We construct a multimodal encoding framework that integrates three complementary feature spaces: 
To address these questions, we first compare human ToM and valence ratings with ratings generated by two recent open-source MLLMs, LLaDA-V~\cite{you2026llada} and Qwen3-VL~\cite{bai2025qwen3}, across three movie stimuli (\emph{The Present}, \emph{Despicable Me}, and \emph{Partly Cloudy}). This allows us to assess both behavioral agreement and the consistency of human--machine neural correspondence across different stimulus contexts. We then examine whether MLLM-derived representations provide information useful for predicting brain activity at the individual-subject level. Our multimodal encoding framework integrates three complementary feature spaces: (1) low-level visual features extracted from a pre-trained deep convolutional network (AlexNet)~\cite{krizhevsky2012imagenet}, (2) language-semantic representations derived from MLLM-generated scene descriptions, and (3) social-cognitive features including ToM, valence, and contextual variables such as the number of on-screen people and face presence. Using banded ridge regression, we decompose the explained variance across feature spaces~\cite{la2022feature} to quantify the contribution of each feature space to regional BOLD responses and compare the roles of perceptual, semantic, and social-cognitive information during naturalistic movie viewing.

We next examined whether children at different ages process the same movie content differently, as reflected in their feature encoding patterns. Naturalistic movie-watching is well suited for studying developmental differences in brain responses. Recent encoding studies have used movie-fMRI to characterize developmental variation in low- and high-level sensory representations in children and adolescents~\cite{mentch2026pregistered}, while voxel-wise encoding work suggests that the neural representation of social information continues to mature from early childhood into adulthood~\cite{im2025early}. We therefore hypothesized that higher-level features, particularly semantic description and social features, would show weaker encoding in younger children than in older children.

The primary purpose of this paper was to evaluate whether MLLM-derived representations can provide meaningful and scalable measures for studying naturalistic brain responses. We focused on their correspondence with human socio-affective judgments, their relationship to neural encoding of perceptual, semantic, and social information, and their variation across development.

\section{Method}

\begin{figure}[!htbp]
    \centering
    \includegraphics[width=\textwidth]{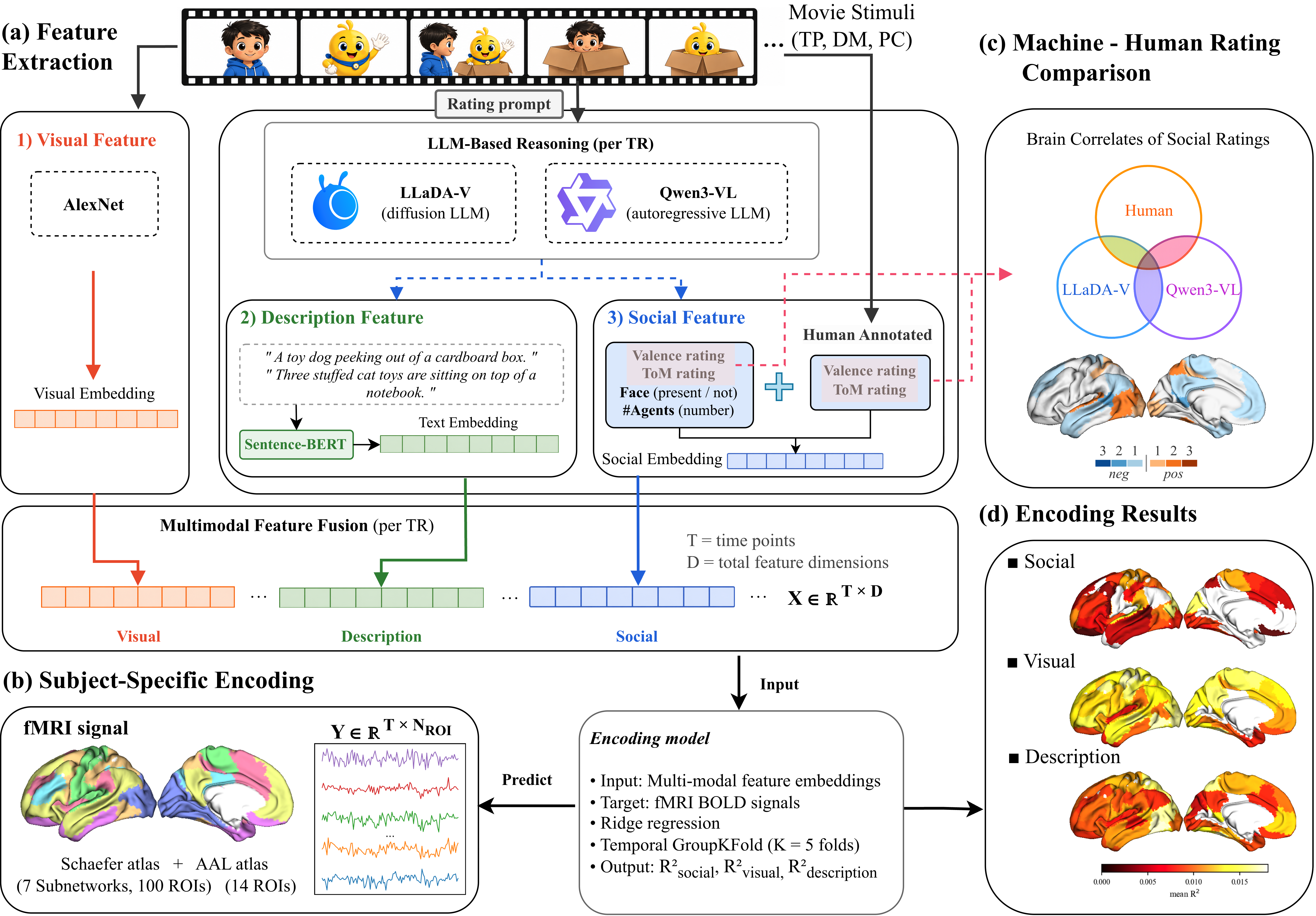}
    \caption{{\bf Overall framework of the proposed multimodal fMRI encoding pipeline.} (a) For each movie frame from the three naturalistic movie-watching datasets (TP, DM, and PC), visual features are extracted using AlexNet~\cite{krizhevsky2012imagenet}, while multimodal LLMs (LLaDA-V~\cite{you2026llada} and Qwen3-VL~\cite{bai2025qwen3}) generate textual descriptions and infer social attributes (e.g., valence, theory-of-mind (ToM), face presence, and the number of people). These feature streams are converted into embeddings and fused into a multimodal representation. (b) The fused features are used to predict subject-specific fMRI BOLD responses using a ridge regression encoding model. (c) LLM-derived social representations are compared with human annotations to assess their correspondence with brain activity. (d) Encoding performance is evaluated for the visual, description, and social feature modalities.}
    \label{fig:pipeline}
\end{figure}

\subsection{Datasets and Participants}

In this study, we utilized previously collected functional magnetic resonance imaging (fMRI) data from three naturalistic movie-watching datasets: {\it The Present (TP), Despicable Me (DM), and Partly Cloudy (PC)}, as shown in Fig.~\ref{fig:pipeline}(a). Specifically, the \textit{TP} and \textit{DM} datasets were sourced from the Healthy Brain Network (HBN) project \cite{alexander2017open}. Participants watched a 3.3-minute clip from \textit{TP} and a 10-minute clip from  \textit{DM}. Following strict quality control, 104 unique participants (61 males and 43 females; aged 5.0 -- 21.9 years, mean $\pm$ SD = 12.0 $\pm$ 4.1 years) were included in the final analysis. Moreover, the \textit{PC} dataset was sourced from the OpenNeuro database~\cite{ds002886:1.1.0}. Participants watched a 5.6-minute silent version of {\it PC}. Following quality control, the final sample consisted of 82 participants.
%divided into \textcolor{orange}{two age groups:} an adult group of 29 participants (17 females and 12 males; aged 18--39 years, mean $\pm$ SD = 24.6 $\pm$ 5.3 years) and a children group of 53 participants (28 females and 25 males; aged 3.5 -- 12.3 years, mean $\pm$ SD = 7.0 $\pm$ 2.5 years).

\subsubsection{fMRI Data Preprocessing and ROI Time-Series Extraction}

We leveraged residual fMRI time series derived from the standard preprocessing pipelines described in our previous work. Specifically, preprocessing of the PC dataset is detailed in~\cite{di2022principal}, while preprocessing of the TP and DM datasets is described in~\cite{di2023individual}.
% All fMRI data underwent a rigorous preprocessing framework. 
Specifically, functional images were realigned, coregistered to each participant's T1-weighted structural image, normalized to standard Montreal Neurological Institute (MNI) space (resampled to 2.4 $\times$ 2.4 $\times$ 2.4 mm$^3$ for TP/DM and 3 $\times$ 3 $\times$ 3 mm$^3$ for PC), and spatially smoothed using an 8 mm Gaussian kernel. Strict head motion quality control was enforced; participants exceeding maximum framewise displacement thresholds (2.4 mm/2.4$^\circ$ for TP/DM; 1.5 mm/1.5$^\circ$ for PC) were excluded. A voxel-wise general linear model (GLM)~\cite{penny2011statistical} was then applied to remove nuisance effects, including Friston's 24 head-motion parameters~\cite{friston1996movement} and low-frequency drifts using a 1/128 Hz high-pass filter. Mean framewise displacement in translation and rotation was also calculated to serve as a covariate in subsequent age-effect analyses.

Following preprocessing, the denoised residual fMRI time series were extracted and temporally aligned with the corresponding movie stimulus features. The final datasets comprised 168 time points for the PC dataset (TR = 2000 ms); 240 time points for the TP dataset (TR = 800 ms; after removing the first 10 time points); and 740 time points for the DM dataset (TR = 800 ms; after removing the first 10 time points). For the TP and DM datasets, the two movie clips were concatenated along the temporal dimension following temporal alignment. The preprocessed voxel-wise fMRI responses were then parcellated into 114 region-of-interest (ROI) time series using two brain atlases: 100 cortical ROIs from the Schaefer atlas~\cite{schaefer2018local} and 14 subcortical ROIs from the Automated Anatomical Labeling (AAL) atlas~\cite{tzourio2002automated}. The resulting time-by-ROI matrices ($Y$) served as the target neural responses for our encoding models, as shown in Fig.~\ref{fig:pipeline}(b).

\subsubsection{Continuous Subjective Ratings}

Continuous subjective ratings were collected from two independent cohorts of healthy participants at the New Jersey Institute of Technology (NJIT) to characterize moment-to-moment fluctuations in perceived movie content. The Theory of Mind (ToM) cohort included 34 participants (17 females; age range: 18--27 years; mean $\pm$ SD: 20.9 $\pm$ 2.3 years), and the affective-valence cohort included 42 participants (8 females; age range: 17--28 years; mean $\pm$ SD: 19.8 $\pm$ 2.3 years). The ToM rating data and procedures have been reported previously~\cite{di2026beyond}. All participants provided IRB-approved informed consent, reported normal or corrected-to-normal sensory function, and had no history of neurological or psychiatric disorders.

In a controlled laboratory setting, participants viewed the same three video clips used in the fMRI experiment. Continuous ratings were collected at 4 Hz using CARMA software~\cite{girard2014carma} and a gaming joystick. The joystick's vertical position was mapped onto a continuous numerical scale ranging from $-10$ to $10$. Before data collection, participants completed a brief practice session.

Participants in the ToM cohort were instructed to continuously rate the extent to which they perceived or inferred the characters' internal mental states (e.g., emotions, thoughts, and intentions), with $-10$ indicating ``not at all'' and $10$ indicating ``very strongly.'' Using the same hardware and numerical scale, participants in the affective-valence cohort continuously rated the moment-to-moment affective valence of each clip, with $-10$ indicating highly negative valence and $10$ indicating highly positive valence.

% Di X, Hanna GB, Biswal BB (2025). Timing Is Everything: Temporal Dynamics of BOLD Responses to Naturalistic Features in Movie Watching. bioRxiv: 10.1101/2025.11.07.687226

\subsection{Multimodal Naturalistic Movie Feature Extraction}

To construct a multi-dimensional feature space for the encoding model, we extracted stimulus representations spanning three complementary modalities from two naturalistic movie stimuli (\textit{TP} and \textit{DM}). The movie (\textit{PC}) was excluded from the encoding analysis because its participant pool did not overlap with those of the other two movies, and its acquisition used a different repetition time, precluding direct concatenation.
Encoding analyses were therefore restricted to the 66 subjects who participated in both \textit{TP} and \textit{DM}. The two movie time series were concatenated along the time axis (TP: $T = 240$\,TRs; DM: $T = 740$\,TRs; combined: $T = 980$\,TRs, where TR\,$= 0.8$\,s) to increase the number of usable timepoints for model estimation.

\subsubsection{Visual Feature Extraction}
Frame-level visual representations were extracted from the full first-convolutional-block (conv1) feature map of a pre-trained AlexNet~\cite{krizhevsky2012imagenet} (64 channels $\times$ 27 $\times$ 27 $=$ 46{,}656 dimensions per frame, rather than a spatially averaged per-channel summary), temporally resampled to the fMRI repetition time (TR = 0.8 s), and reduced to 128 dimensions using principal component analysis (PCA), fit jointly across the concatenated TP and DM movie frames. In the beginning, we summarized conv1 by spatially averaging each channel, yielding only a 64-dimensional per-frame vector. This pooled representation discards the spatial layout of the feature map and collapses all within-channel variation to a single mean, so most of the low-level spatial structure (edges, local contrast, and their retinotopic arrangement) that early visual cortex is sensitive to is lost. Consistent with this, the pooled 64-dimensional features explained little variance in visual regions. We therefore retained the full spatially resolved conv1 activation and used joint PCA for dimensionality reduction, which preserved this spatial information and substantially improved the visual band's contribution to the encoding model.

\subsubsection{Semantic Description Feature Extraction}
Scene-level textual descriptions were generated at each fMRI repetition time (TR\,$= 0.8$\,s) using LLaDA-V (Large Language Diffusion Model with Vision)~\cite{you2026llada}, a multimodal diffusion language model. For each sampled video frame, LLaDA-V was prompted to produce a free-form natural-language description of the visual scene content. These frame-level descriptions were then encoded into 768-dimensional dense sentence embeddings using sBERT~\cite{reimers2019sentence} (all-mpnet-base-v2, huggingface.co), a sentence encoder built on the MPNet transformer~\cite{song2020mpnet} and trained with a contrastive (siamese) objective to produce embeddings in which semantically similar sentences lie close together.
Prior to model fitting, the embeddings from the two movies were concatenated, z-scored, and reduced to 64 dimensions using principal component analysis (PCA) fit jointly across both movies. The resulting 64-dimensional time series were then temporally smoothed with a Gaussian kernel ($\sigma = 2$\,s, i.e.\ $\approx 2.5$\,TR at TR $= 0.8$\,s) to attenuate high-frequency noise and better match the temporal scale of the BOLD signal.

\subsubsection{Social-Cognitive Feature Extraction}
The social feature vector comprised six continuous time series at each fMRI repetition time: two machine-generated ratings (ToM and valence), two human behavioral ratings (ToM and valence), and two low-level social features (number of on-screen people and face presence, binary).

{\bf Machine-generated ratings.}
Our prior study~\cite{huang2026brain} systematically compared ToM and valence ratings generated by three multimodal large language models, LLaDA-V~\cite{you2026llada}, Qwen3-VL~\cite{bai2025qwen3}, and Gemma 3~\cite{team2025gemma}, with human behavioral annotations and included detailed ablation analyses of context window length. Based on these comparisons, we selected LLaDA-V and Qwen3-VL for the present study because they showed more consistent alignment with human ratings, whereas Gemma 3 exhibited a negative correlation with human ToM ratings for \textit{Despicable Me}. Although our prior study evaluated configurations with varying amounts of preceding context, all machine-generated ratings analyzed in the present study were obtained using the base (no-context) configuration, in which each video frame was evaluated independently without preceding contextual descriptions. Both models were included in the rating comparisons and analyses. Only for the subsequent fMRI encoding analyses did we select the model whose base ratings showed the highest correlation with human annotations for each movie: LLaDA-V for \textit{The Present} and Qwen3-VL for \textit{Despicable Me}. Because the machine-generated ratings exhibited more rapid frame-to-frame fluctuations than the continuously collected human ratings, after alignment with the fMRI TR grid we applied a 20-TR moving-average filter ($16\,\mathrm{s}$ at TR $= 0.8\,\mathrm{s}$), separately within each movie, to reduce these fluctuations.

\textbf{Human ratings.}
Behavioral annotations of ToM and valence were collected from groups of independent raters. A leave-one-out (LOO) reliability procedure was applied, and raters falling in the bottom 20\% of inter-rater agreement were excluded.
The resulting mean time series were resampled to the fMRI TR grid via linear interpolation without additional smoothing.

\textbf{Low-level social features.}
The number of on-screen people and a binary face-presence indicator were extracted TR-by-TR using LLaDA-V. All six scalar time series were $z$-scored and concatenated to form the final \textit{social} feature vector.

\subsection{Human and Machine Ratings Comparison}

Detailed comparisons between machine-generated and human ratings are provided in our prior study~\cite{huang2026brain}. Here, we visualize their temporal correspondence using the raw human and Qwen3-VL valence rating time series for \textit{Despicable Me}, with machine ratings generated using the base (single-frame, no-context) prompt. See Appendix~\ref{app:machine_prompt} for detailed prompts.

\subsection{Univariate BOLD correlates with Human and Machine Ratings}

Because machine and human ratings matched so closely, we next examined whether human and machine ratings were associated with activity in overlapping brain regions.

\subsubsection{Movie-Specific ROI Correlation Analysis}
% overlap between human and machine / detect similarity among human and machine 

To characterize the movie-specific neural correlates of ToM and valence, we applied the same ROI-wise correlation pipeline independently to each movie (\emph{The Present}, \emph{Despicable Me}, \emph{Partly Cloudy}). For each movie, rating source (human, LLaDA-V, Qwen3-VL), and measure, the fMRI signal was shifted relative to the rating time series by a fixed hemodynamic lag, Pearson correlations between the rating and each ROI's response were then computed per subject, Fisher-$z$-transformed, and tested against zero with a one-sample $t$-test (FDR-BH corrected, $\alpha = 0.05$). Because the temporal alignment between naturalistic features and BOLD responses can vary across feature types, particularly for subjective ratings~\cite{di2025timing}, we estimated a separate fixed lag for each measure. The fixed lag was selected by sweeping candidate lags from $0.0$ to $6.4$\,s (in $0.8$\,s steps) and choosing the value that maximized the number of ROIs in which model and human ratings were both FDR-significant with the same correlation sign in all three movies. Lags were converted to the nearest whole TR within each movie. For TP and DM, the nominal lags were 3.2 s for ToM and 4.8 s for valence. For PC, rounding these lags to the nearest whole TR resulted in an effective lag of 4.0 s for both measures.

Results are displayed as bar plots of mean $\pm$ SEM Pearson $r$ per ROI, grouped by network, with model-generated and human ratings shown side by side. For each movie we display only ROIs that (i) survived FDR correction for both the model and the human ratings and (ii) shared the same sign of correlation in both; ROIs where the model and human disagreed in direction were excluded. 

% See Figures~\ref{fig:tom_bar2}, \ref{fig:val_bar2} and~\ref{fig:brain2}.
To visualize where the model--human overlap was reproducible across movies, we combined the per-movie ROI-wise correlations described above into a single consistency map for each measure (ToM, valence) and each model (LLaDA-V, Qwen3-VL). For every ROI, we counted the number of movies (out of three) in which the model and human ratings were both FDR-significant ($q < 0.05$) and correlated with that ROI in the same direction. Each ROI was then assigned (i) this movie count (1, 2, or 3), rendered as increasing color depth, and (ii) the sign of its correlation averaged over the qualifying movies, rendered as hue (orange for positive, blue for negative); ROIs that did not meet the criterion in any movie were left uncolored. Maps were projected onto the fsLR (Conte69) surface using the native Schaefer2018 100-parcel, 7-network parcellation in the \texttt{BrainSpace} library~\cite{vos2020brainspace, schaefer2018local}.

\subsubsection{Cross-Movie ROI Correlation Analysis}

To identify brain regions whose responses to ToM and valence are consistent across movies and rating sources, we performed a cross-movie analysis at the fixed lags described above.
For each movie (\textit{The Present}, \textit{Despicable Me}, \textit{Partly Cloudy}), rating source (human, LLaDA-V, Qwen3-VL), and measure, the fMRI signal of each ROI was shifted forward by the selected lag, and Pearson correlations between each ROI time series and the rating signal were computed per subject, Fisher-$z$-transformed, and tested against zero across subjects with a one-sample $t$-test; $p$-values were FDR-corrected across ROIs (Benjamini--Hochberg, $\alpha = 0.05$).

We first characterized each rating source separately: an ROI was retained if it was FDR-significant in all three movies and had the same sign of mean correlation across movies. For these cross-movie-consistent ROIs we display the movie-wise correlations (mean $\pm$ SEM) as bar plots (Figure~\ref{fig:bar1}). We then quantified the overlap between rating sources: an ROI was counted as shared by a pair of sources if it was cross-movie-consistent for both sources and had matching correlation directions between them. For ToM we report the human--model overlap (human $\cap$ LLaDA-V and human $\cap$ Qwen3-VL). For valence, the human ratings yielded only a single cross-movie-consistent ROI. We therefore report the valence overlap in the text and restrict the cortical-surface visualization to ToM. The overlapping ROIs are shown on the cortical surface, colored by the average Pearson correlation across the contributing movies and sources (Figure~\ref{fig:brain1}).

\subsection{Encoding Model}
\subsubsection{Feature-Space Similarity Analysis}

To characterize the pairwise relationships among the extracted feature spaces, we computed a similarity matrix over all the raw (unsmoothed) feature pairs for each movie. The similarity measure was adapted to the dimensionality of each pair. 
For one-dimensional feature pairs (e.g.\ scalar social ratings), similarity was quantified as the Pearson correlation evaluated on held-out data using a
3-fold \texttt{GroupKFold} cross-validation, with non-overlapping temporal windows of 15\,TRs as grouping units to account for temporal autocorrelation (8\,TRs for \textit{Partly Cloudy}, which has a longer TR of 2\,s).

For higher-dimensional feature pairs, we employed $\ell_2$-regularized Canonical Correlation Analysis (rCCA)~\cite{chapman2021cca} via the \texttt{CCA-Zoo}
library, extracting the first canonical dimension ($n_\text{components}=1$). Regularization was selected via nested cross-validation: an inner 3-fold \texttt{GroupKFold} loop searched over 11 candidate values ($c \in \{0\} \cup [10^{-5}, 10^{0}]$, log-spaced), and the value maximizing the mean held-out canonical correlation was used to refit the model on the full outer-training set. The similarity score for each pair was defined as the Pearson correlation between the first canonical variates of the two feature spaces on the outer test set, averaged across the 3 outer folds. %The resulting pairwise similarity matrices are reported separately for each movie (Figure~\ref{fig:corr}).

\subsubsection{ROI-wise fMRI Encoding Model}
We fit a separate encoding model for each subject, predicting the BOLD time course of each ROI from stimulus-derived feature spaces. Successful prediction indicates that information captured by a feature space is reflected in the regional neural response~\cite{naselaris2011encoding}. This approach adapts the voxel-wise encoding framework~\cite{dupre2025voxelwise} to ROI-level responses rather than individual voxels. We used \textbf{banded ridge regression}~\cite{nunez2019voxelwise}, implemented with the \texttt{himalaya} library~\cite{la2022feature}, to jointly model the three feature spaces while allowing each band (social, description, and visual) to have its own regularization weight.
To account for the delayed and dispersed hemodynamic response, we followed previous work~\cite{huth2016natural, deniz2023semantic} and replaced each feature space with three time-delayed copies of itself, concatenated along the feature dimension, at delays of 4, 6 and 8 TRs (3.2, 4.8 and 6.4~s at TR~$=0.8$~s), so that the model learns a separate weight for each delay.
Because the total feature dimensionality after delay expansion (594) was smaller than the number of training time points (approximately 770 per outer fold), we used the primal \texttt{BandedRidgeCV} formulation rather than the kernelized \texttt{MultipleKernelRidgeCV}. For linear kernels, multiple-kernel ridge regression is equivalent to group-regularized (banded) ridge regression, while the primal formulation is computationally more efficient when the feature dimension does not exceed the number of samples~\cite{la2022feature}.
Features were z-scored using training-fold statistics, which were then applied to the test fold; fMRI targets were z-scored separately within the training and test sets of each fold. Because the TP and DM movies were concatenated within each participant, we additionally included a run indicator (coded $+1$ for TP and $-1$ for DM) as a separate band to absorb any mean signal difference between the two movies. This nuisance regressor was set to zero on the held-out fold, so that it never contributed to the test-set predictions or $R^{2}$, and it was excluded from the product-measure decomposition of per-feature contributions.

Hyperparameters were selected over a grid of 40 log-spaced regularization values ($\alpha \in [10^{-3}, 10^{10}]$) using a random-search solver (500 iterations, SVD diagonalization). The model was fit using nested cross-validation with 5 outer and 5 inner folds, both implemented as \texttt{GroupKFold} over non-overlapping, contiguous temporal chunks, so that all TRs within a chunk were assigned to the same fold. Inner folds were drawn only from the outer training chunks. We evaluated chunk lengths of 10, 12, 16 and 20 TRs, using 20 TRs in the main analysis. When the two movies were concatenated, chunks were defined separately within each movie so that no chunk straddled the movie boundary, and the FIR delays were likewise applied separately within each movie. The contribution of each feature space to the predicted BOLD response was quantified with the product-measure decomposition (\texttt{r2\_score\_split}), which partitions the joint $R^{2}$ of each brain region into per-feature components that sum to the joint $R^{2}$.

The pipeline was applied to 66 participants, each encoded individually, across 114 brain regions: 100 cortical parcels from the Schaefer 2018 atlas (7-network version)~\cite{schaefer2018local} and 14 subcortical and medial temporal regions from the AAL atlas~\cite{tzourio2002automated} (bilateral hippocampus, parahippocampal gyrus, amygdala, caudate, putamen, pallidum and thalamus). The final feature set comprised three groups: \textit{social}, \textit{description} (sBERT), and \textit{visual} (AlexNet \texttt{conv1}).

\subsubsection{Age Effects}

Participants were divided into three equal-sized age tertiles by sorting all 66 participants by age: young ($n=22$; $5.2$--$10.3$~yr, mean $7.8$), middle ($n=22$; $10.4$--$14.4$~yr, mean $12.4$), and old ($n=22$; $14.5$--$21.9$~yr, mean $17.4$); per-participant product-measure $R^2$ values were averaged across the ROIs of each cortical network, and every feature $\times$ network combination was tested for age effects with all three pairwise tertile comparisons (young vs.\ middle, middle vs.\ old, young vs.\ old) using a two-tailed label-permutation test (10{,}000 permutations, add-one smoothing) thresholded at an uncorrected $p<0.05$.

\section{Results}

\subsection{Human and Machine Rating Similarity}

\subsubsection{Machine Ratings Track Human Ratings}
\label{sec:result1}

\begin{figure}[!htbp]
    \centering
    \includegraphics[width=0.89\textwidth]{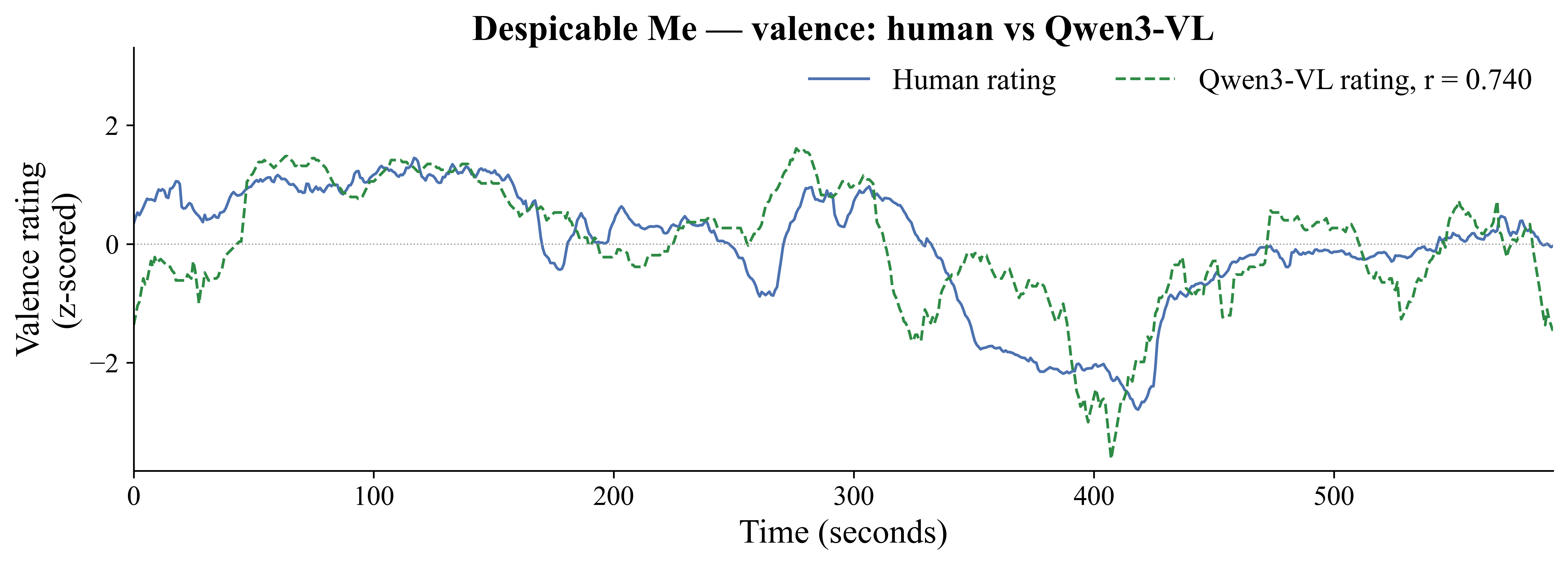}
    \caption{\textbf{Machine and human valence ratings for \textit{Despicable Me}.} Group-mean human valence rating (blue) and Qwen3-VL valence rating (green, dashed) on the fMRI time grid (TR = 0.8\,s), starting 8\,s after movie onset. The model rating in this figure was smoothed with a 20\,s moving average; the human rating was linearly interpolated and not further smoothed. Both time courses were z-scored to compare their trends rather than their amplitudes ($r = 0.740$, zero lag).}
    \label{fig:thrating}
\end{figure}

A systematic comparison of machine and human ratings, covering all models, movies, rating dimensions and prompting strategies, is reported in our previous work~\cite{huang2026brain}. Here we focus on whether the machine ratings follow the moment-to-moment trajectory of human judgments. Figure~\ref{fig:thrating} overlays the human and Qwen3-VL valence time courses for \textit{Despicable Me}. The machine rating closely tracks the rises and falls of the human rating over the full movie ($r = 0.740$). Agreement of this kind nevertheless depends on the stimulus and the rating dimension~\cite{huang2026brain}. We therefore treat MLLM ratings as a stimulus-dependent complement to human annotation, not as interchangeable with human judgments.

% 
% delete a

\subsubsection{Movie-Specific Neural Correlate Overlap}
% \subsubsection{Overlapping neural correlates within each movie}

\begin{figure}[!htbp]
    \centering
    \includegraphics[width=\textwidth]{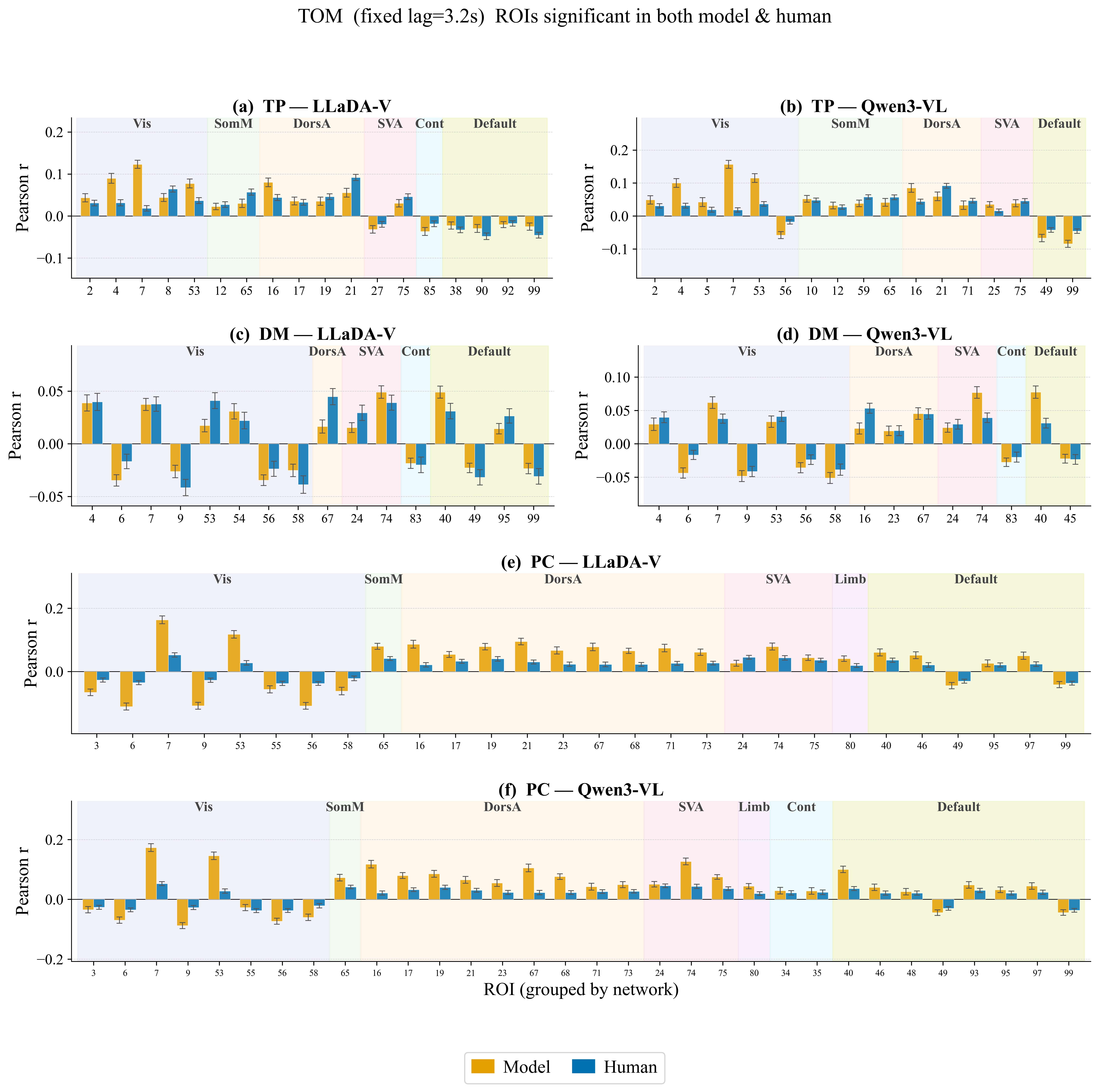}
    \caption{\textbf{ToM: model vs.\ human rating correlations with BOLD.} (fixed lag: 3.2~s for TP/DM and 4.0~s for PC). Mean per-ROI Pearson correlation ($\pm$SEM across subjects) of the model rating (orange) and human rating (blue) with fMRI activity. Panels (a--f) span the three movies $\times$ two models (LLaDA-V, Qwen3-VL). Only ROIs significant for both model and human (FDR $q < 0.05$) and of the same sign are shown, grouped by network; panel width scales with ROI count.}

    \label{fig:tom_bar2}
\end{figure}

\begin{figure}[!htbp]
    \centering
    \includegraphics[width=\textwidth]{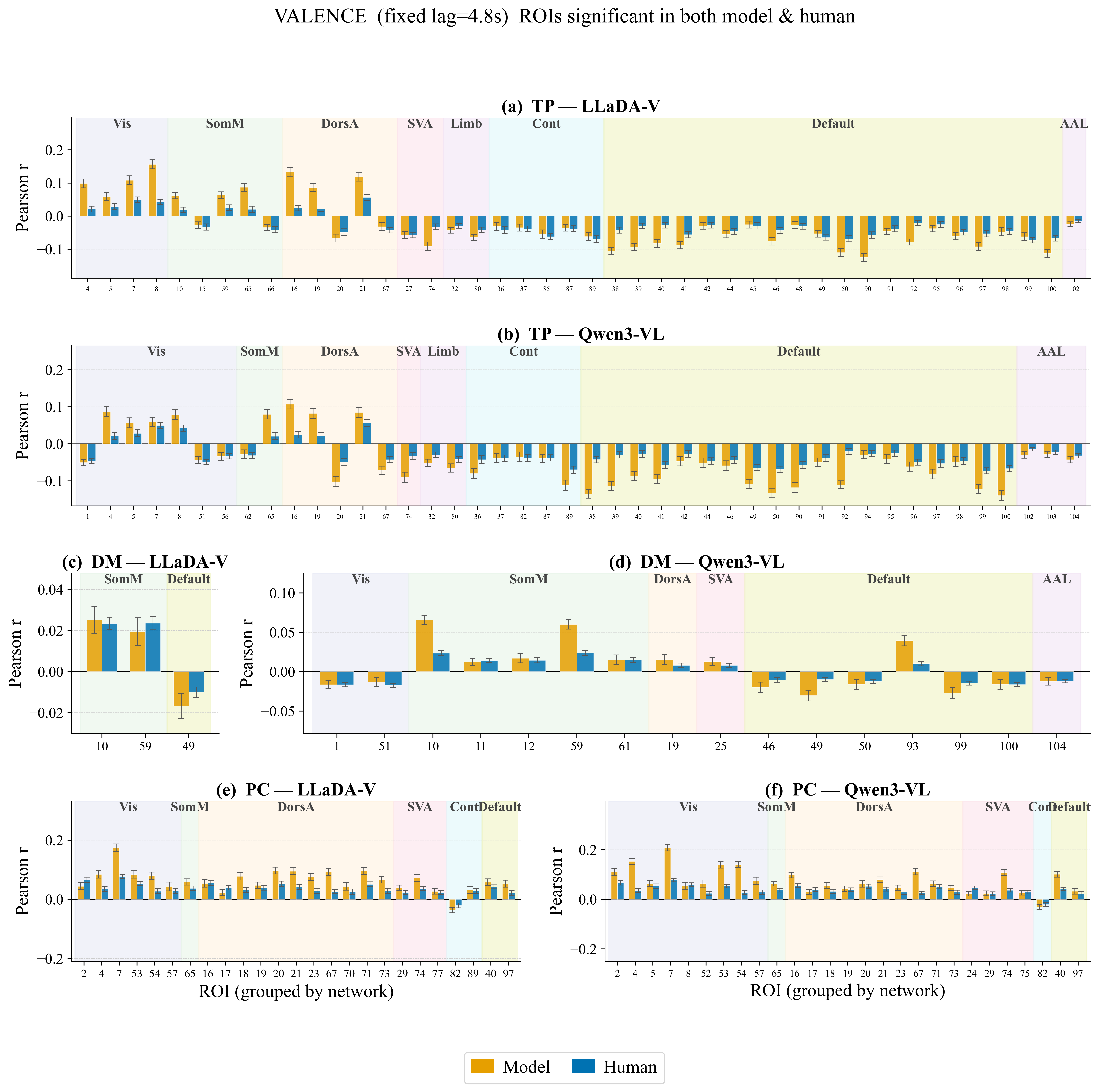}
    \caption{\textbf{Valence: model vs.\ human rating correlates with BOLD.} (fixed lag: 4.8~s for TP/DM and 4.0~s for PC). Mean per-ROI Pearson correlation ($\pm$SEM) of model (orange) and human (blue) ratings with fMRI activity; panels (a--f) cover the three movies $\times$ two models. ROIs are significant for both model and human (FDR $q < 0.05$), same-signed, and grouped by network, with panel width proportional to ROI count.}

    \label{fig:val_bar2}
\end{figure}

\begin{figure}[!htbp]
    \centering
    \includegraphics[width=\textwidth]{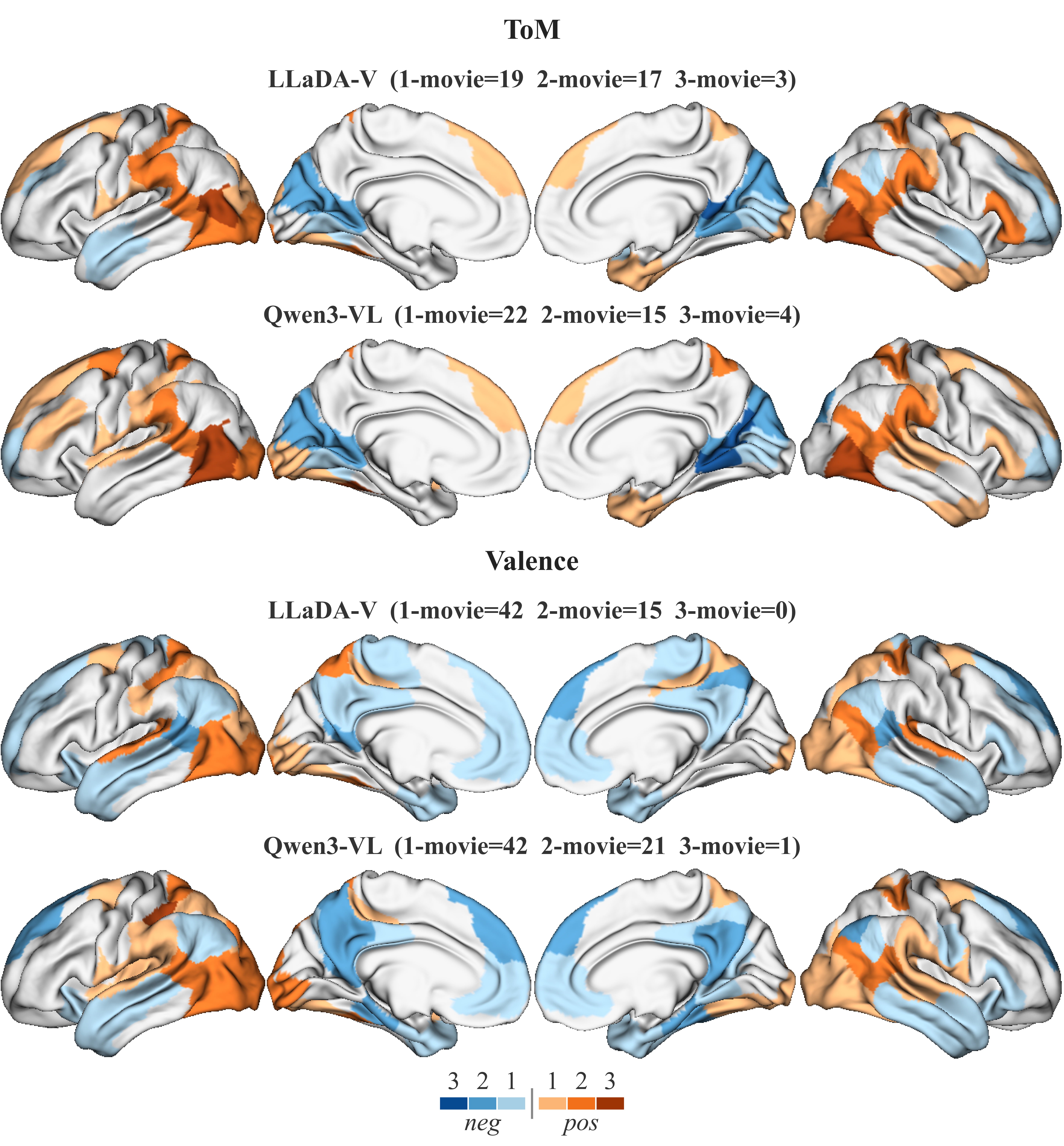}
    \caption{\textbf{Cross-movie consistency of model--human ROIs for ToM and valence.} Each ROI is coloured by the number of movies (1--3; light$\rightarrow$dark) in which the model and human ratings both correlate significantly (FDR $q < 0.05$) with fMRI activity and share the same sign; orange = positive, blue = negative. Top two rows: ToM (fixed lag: 3.2~s for TP/DM and 4.0~s for PC); bottom two rows: valence (fixed lag: 4.8~s for TP/DM and 4.0~s for PC). Within each measure, rows correspond to LLaDA-V and Qwen3-VL.}

    \label{fig:brain2}
\end{figure}

To directly compare model- and human-derived signals, we plotted the ROI-wise correlations (mean $\pm$ SEM across subjects) for every ROI that was significant for both the model and the human ratings within a given movie, with the two rating sources shown side by side and ROIs grouped by network (Figures~\ref{fig:tom_bar2}, \ref{fig:val_bar2}).

For ToM, the jointly significant correlations were predominantly positive and concentrated in the visual and dorsal-attention networks, with additional effects in salience/ventral-attention regions. Model and human correlations shared the same direction throughout (by construction). The number of jointly significant ROIs varied across movies, being largest for \textit{Partly Cloudy} (28 for LLaDA-V, 32 for Qwen3-VL), where a broad set of dorsal-attention ROIs was engaged, and smallest for \textit{Despicable Me} (16 and 15). Model correlations exceeded the human ones in magnitude most clearly in \textit{Partly Cloudy}; in \textit{The Present} this held for Qwen3-VL but not LLaDA-V, and in \textit{Despicable Me} the two were of comparable size.

For valence, the pattern was strongly movie-dependent. In \textit{The Present}, jointly significant ROIs were dominated by negative correlations across the default-mode network, alongside positive correlations in early visual and dorsal-attention regions. In \textit{Partly Cloudy}, by contrast, nearly all jointly significant correlations were positive, concentrated in dorsal-attention and visual regions. \textit{Despicable Me} yielded few jointly significant ROIs, particularly for LLaDA-V (3, versus 16 for Qwen3-VL). Model correlations exceeded the human ones in magnitude in \textit{The Present} and \textit{Partly Cloudy}, whereas in \textit{Despicable Me} both were weak and of comparable size.

Figure~\ref{fig:brain2} maps the model--human overlap across movies on the cortical surface.
Across-movie consistency differed markedly between the two measures. For ToM, both models showed a small set of ROIs with model–human overlap in all three movies. LLaDA-V showed overlap in 19, 17, and 3 ROIs for one, two, and three movies, respectively, and Qwen3-VL in 22, 15, and 4. Compared with LLaDA-V, Qwen3-VL had more single-movie ROIs and fewer two-movie ROIs, suggesting somewhat more movie-specific effects. For valence, model–human overlap was more frequent within individual movies but rarely generalized across them. LLaDA-V showed overlap in 42, 15, and 0 ROIs for one, two, and three movies, respectively, and Qwen3-VL in 42, 21, and 1. The only ROI consistent across all three movies was a left posterior dorsal-attention parcel (LH\_DorsAttn\_Post\_4), which reappears in the cross-movie analysis below (Section~\ref{sec:cross}). Thus, valence correspondence was more stimulus-dependent than ToM correspondence. The valence maps were more dominated by negative correlations than the ToM maps, with broad blue regions across the default-mode network (medial prefrontal, posterior cingulate/precuneus, and lateral temporal cortex) indicating ROIs whose activity was associated with more negative affective moments. These negative effects were driven largely by \textit{The Present}. In contrast, ToM correlations were predominantly positive, with negative correlations confined mainly to medial occipital visual cortex.

\subsubsection{Cross-Movie Neural Correlate Consistency}
\label{sec:cross}
% \subsubsection{Human–machine overlap consistent across movies}

\begin{figure}[!htbp]
    \centering
    \includegraphics[width=\textwidth, trim = {0 0 0 15mm}, clip]{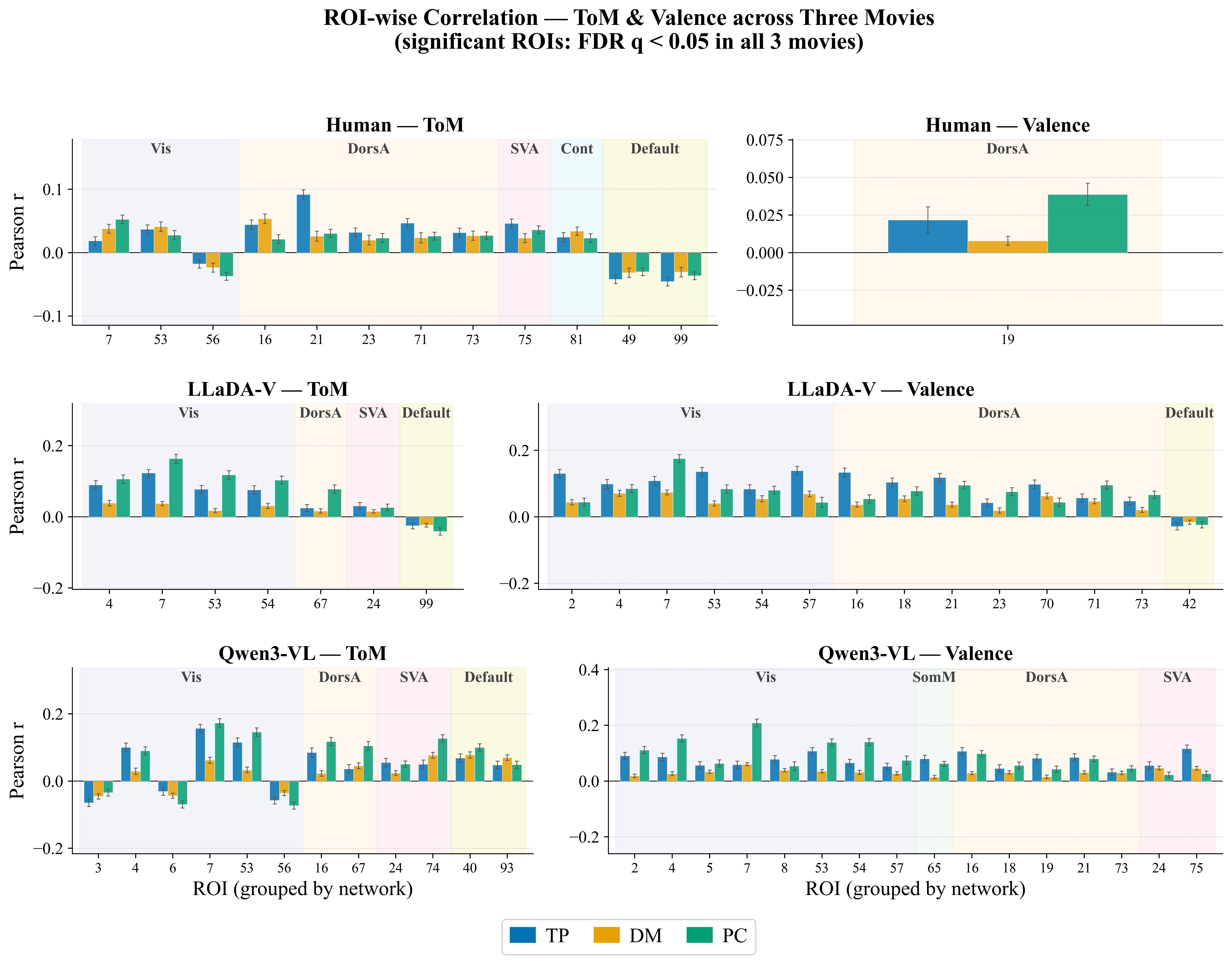}
    \caption{\textbf{ROI-wise correlation with ratings across three movies.} Mean per-ROI Pearson correlation (bars, $\pm$SEM across subjects) between each rating source (rows: Human, LLaDA-V, Qwen3-VL) and fMRI activity, for ToM and valence (columns), shown separately for the three movies (TP, DM, PC). Only ROIs significant in all three movies (FDR $q < 0.05$) and sharing the same sign across movies are plotted; ROIs are grouped by network (shading), and panel width scales with ROI count.}

    \label{fig:bar1}
\end{figure}

% legend 
% SEM
\begin{figure}[!htbp]
    \centering
    \includegraphics[width=\textwidth]{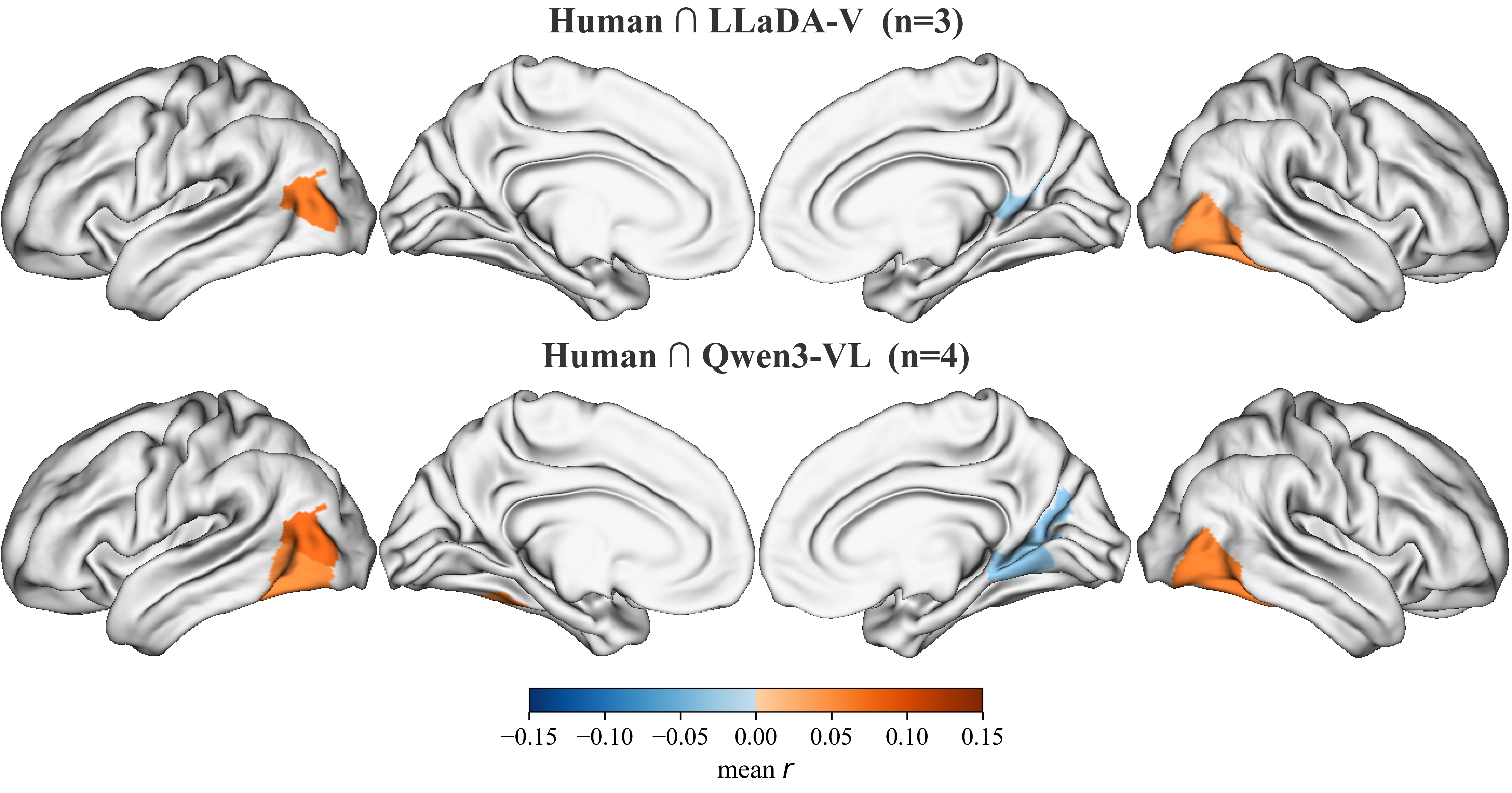}
    \caption{\textbf{ToM: cortical ROIs shared by model and human ratings.} (fixed lag: 3.2~s for TP/DM and 4.0~s for PC). ROIs where both the model and the human ToM rating correlate significantly with fMRI activity (FDR $q < 0.05$) in \emph{all three} movies, with the same sign across movies, for the human$\cap$LLaDA-V and human$\cap$Qwen3-VL overlaps. Colour gives the mean correlation averaged across movies (orange = positive, blue = negative). }

    \label{fig:brain1}
\end{figure}

To identify brain regions whose rating--BOLD relationship is stable across movies, we display, for each rating source and measure, the ROIs that were significant across all three movies (Figure~\ref{fig:bar1}). Pearson correlations between each ROI's fMRI response and the annotation time series are shown for ToM (left) and valence (right), separately for human ratings (top), LLaDA-V (middle), and Qwen3-VL (bottom). Only ROIs surviving FDR correction ($q < 0.05$) in all three movies are shown, grouped by functional network. Each bar is the cross-subject mean correlation ($\pm$ SEM across subjects) for three movies, using fixed lags of 3.2~s for ToM and 4.8~s for valence for TP/DM (4.0 s for PC). An ROI was retained only if it was FDR-significant and shared the same correlation sign across all three movies. 
% The human--model overlap additionally required matching correlation direction between the two sources.

For ToM, 12 ROIs were cross-movie-consistent for human ratings, 7 for LLaDA-V, and 12 for Qwen3-VL. Of these, three were shared by Human~$\cap$~LLaDA-V and four by Human~$\cap$~Qwen3-VL (Figure~\ref{fig:brain1}). Both overlaps included the same two positively correlated parcels in bilateral lateral occipitotemporal cortex (LH\_Vis\_7 and RH\_Vis\_3). Human~$\cap$~LLaDA-V additionally included a ventral posterior cingulate parcel of the default network (RH\_Default\_pCunPCC\_1), which correlated negatively with both human and LLaDA-V ratings in all three movies. Human~$\cap$~Qwen3-VL additionally included a positively correlated left lateral occipitotemporal parcel assigned to the dorsal-attention network (LH\_DorsAttn\_Post\_1) and a negatively correlated parcel in medial occipital cortex (RH\_Vis\_6). Thus, a small set of ROIs showed stable human--model correspondence for ToM across all three movies.

For valence, human ratings were far less consistent across movies than model ratings. Only one ROI was FDR-significant with the same correlation sign in all three movies for human ratings. This ROI, a left anterior intraparietal parcel of the dorsal-attention network (LH\_DorsAttn\_Post\_4), was also positively correlated with Qwen3-VL ratings in all three movies, whereas no ROI was shared with LLaDA-V. This sparse overlap suggests that the neural correlates of human valence ratings depended strongly on stimulus context. We therefore did not display a valence overlap map.

\subsection{MLLM-derived features predict brain activity and capture its development}

\subsubsection{Similarity Across Feature Spaces}

\begin{figure}[!htbp]
    \centering

    \centering
    \includegraphics[width=\textwidth]{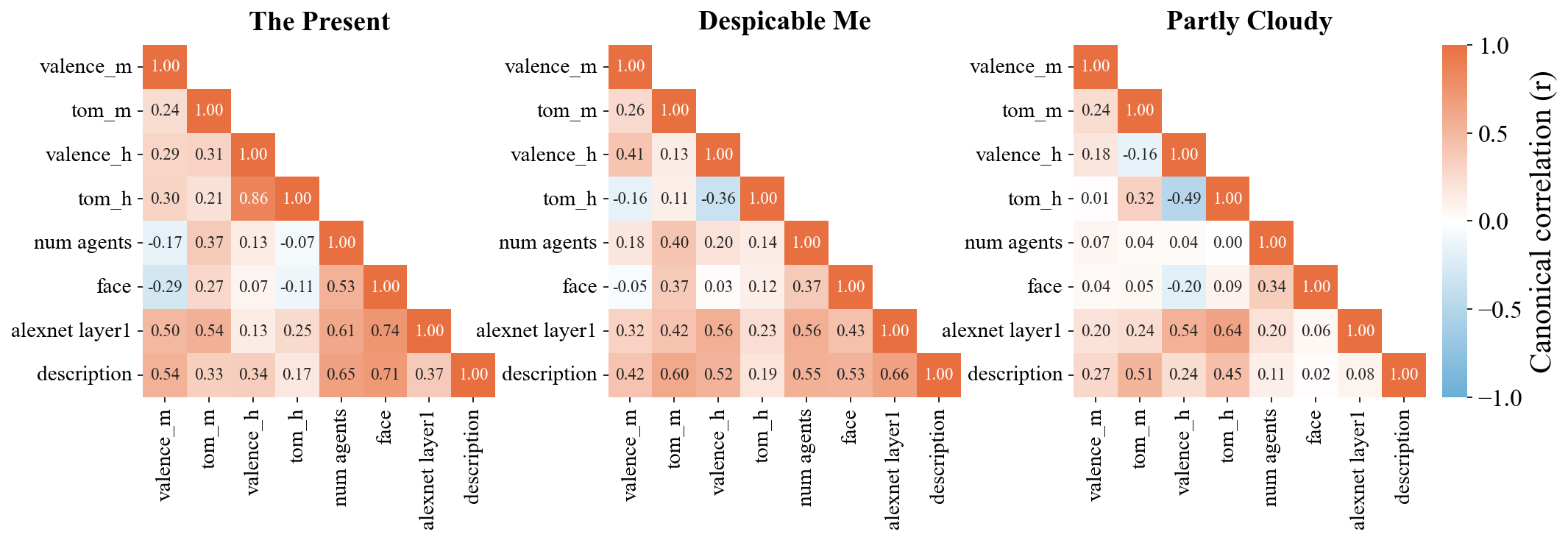}
    \caption{Pairwise similarity among stimulus features in the three movies. Scalar feature pairs were compared with Pearson correlation; pairs involving multidimensional features (AlexNet layer~1, description embedding) were compared with cross-validated regularized CCA. Subscripts m and h denote MLLM and human ratings, respectively.}
    \label{fig:feature_similarity}

    \label{fig:corr}

\end{figure}

Feature correlations showed clear movie-specific structure. See Figure \ref{fig:corr}. In The Present and Despicable Me, low-level social features were strongly coupled with visual and description features: number of people and face presence showed high correlations with AlexNet layer1 and description embeddings. In contrast, these relationships were much weaker in Partly Cloudy, where number of people and face presence were only weakly related to ToM, valence, visual features, and description embeddings.

Across movies, ToM and valence showed moderate correlations with visual and description features, but the strength of these relationships varied by movie. Description embeddings were especially correlated with ToM in Despicable Me and Partly Cloudy, suggesting that the descriptions captured high-level social and mental-state information. In contrast, the correlation between AlexNet layer1 and description was high in Despicable Me but near zero in Partly Cloudy, indicating that the description feature captured information beyond low-level visual structure in this movie.

The relationship between human ToM and human valence varied substantially across movies, being strongly positive in The Present but negative in Despicable Me and Partly Cloudy. This suggests that the coupling between affective and mental-state information depends strongly on narrative context.

% Overall, the results revealed that low-level social features exhibited modest correlations with visual features, reflecting shared sensitivity to scene composition (e.g., number of people and face presence). In contrast, higher-level social-cognitive features such as ToM showed weaker alignment with visual representations but stronger associations with description-based features, suggesting a closer link to semantic and narrative content.

% CCA analysis further indicated that the social and description feature spaces shared substantial common variance, whereas visual features were comparatively more distinct. Together, these findings suggest that social-cognitive representations occupy an intermediate position, bridging perceptual inputs and higher-level semantic descriptions.

% \subsection{Implementation Details}

% Each feature space was standardized independently using z-scoring based on the training data within each cross-validation fold. The same transformation was then applied to the held-out test data to prevent information leakage.

\subsubsection{Vision and description features predict brain activity}

We fitted a separate model for each participant using the concatenated TP and DM time series. For each of the three feature spaces (social, visual, and description), we used the product-measure decomposition to quantify its contribution to the joint model’s predictive R².

\begin{figure}[!htbp]
    \centering
    \includegraphics[
    width=\textwidth, 
    trim={0 0 0 215},
    clip]{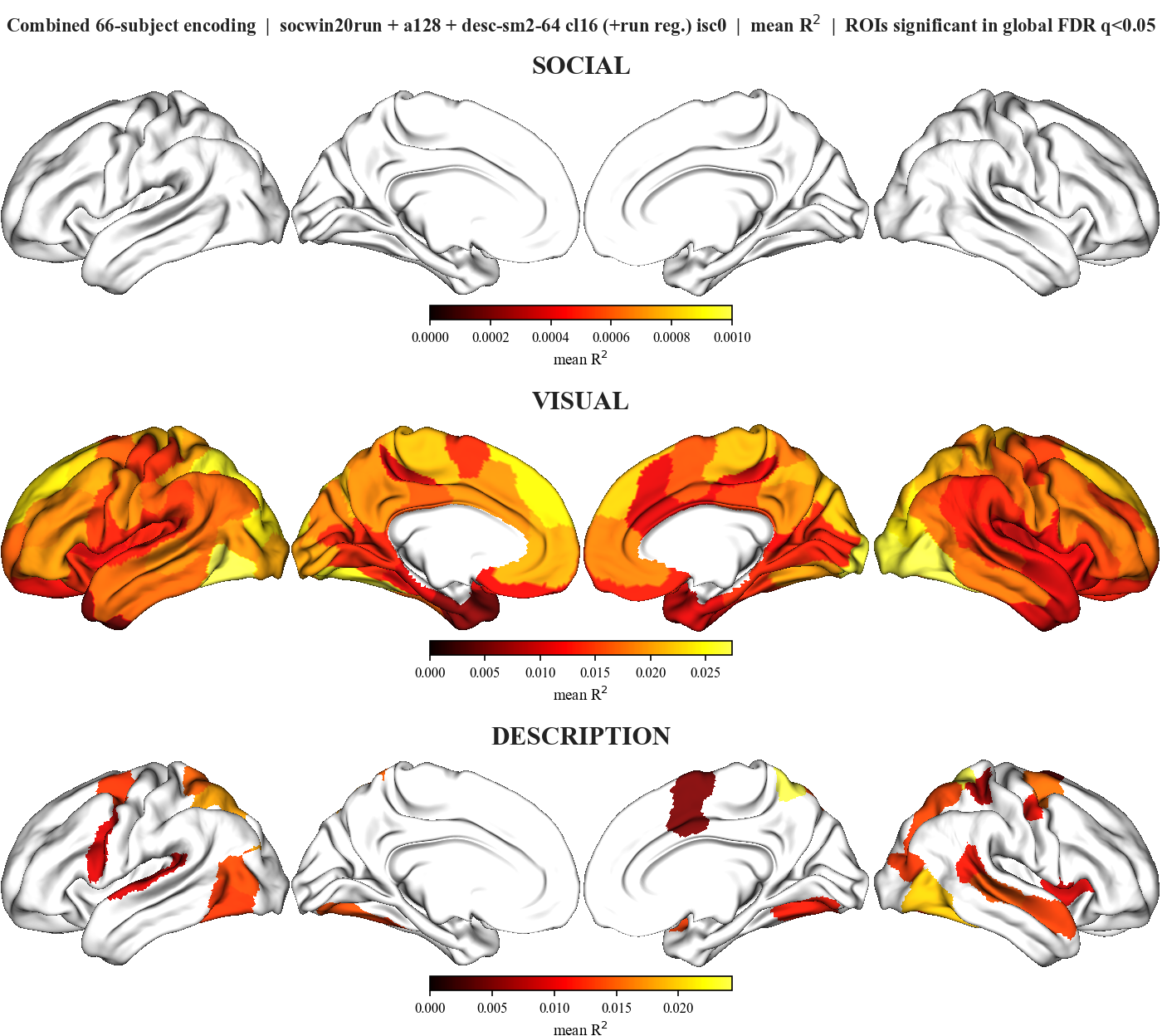}
    \caption{\textbf{Cortical surface maps of encoding performance for the visual (top) and description (bottom) features.} Encoding performance is quantified by the mean coefficient of determination ($R^{2}$), averaged across the full sample of 66 participants, and projected onto the Conte69 cortical surface with the \texttt{BrainSpace} library \citep{vos2020brainspace}. Parcel-wise $R^{2}$ values were computed on the Schaefer2018 100-parcel, 7-network atlas \citep{schaefer2018local} in its native fsLR (Conte69) surface space and assigned to surface vertices using BrainSpace's \texttt{map\_to\_labels} \citep{vos2020brainspace}.
    Colours indicate the parcel-wise mean $R^{2}$. Only parcels with significantly positive encoding after FDR correction (one-sided Wilcoxon signed-rank test against zero across the 66 participants, Benjamini--Hochberg $q < 0.05$) are coloured; non-significant parcels are shown in white. The \emph{social} feature is not shown because it did not survive FDR correction in any cortical parcel and its mean $R^{2}$ was negative throughout. Subcortical and cerebellar regions are not displayed. Visual feature dimension is 128 and description feature dimension is 64, chunk length is 16.}
    \label{fig:encoding_map}
\end{figure}

Cortical surface maps (Figure~\ref{fig:encoding_map}) illustrate the spatial distribution of encoding performance across ROIs for all 66 participants combined. Low-level visual features (AlexNet conv-1) showed widespread predictive performance across cortex, whereas language-description encoding was more spatially selective. The social feature did not yield any ROIs that survived multiple-comparison correction. 
Visual encoding was strongest in visual and dorsal-attention networks but was broadly distributed, remaining comparably high in default-mode and control networks. 
Description encoding was more spatially selective and was concentrated in the dorsal-attention network, spanning superior parietal and intraparietal cortex, lateral occipitotemporal cortex, and the bilateral frontal eye fields. Additional significant parcels lay in lateral and ventral occipital visual cortex, in right posterior superior temporal sulcus extending to the temporo-parietal junction (RH\_Default\_Temp\_3 and RH\_SalVentAttn\_TempOccPar\_1), in right anterior temporal cortex and right anterior insula, in somatomotor cortex, and in the left parahippocampal gyrus. This pattern suggests that scene descriptions captured distributed visual and attentional information, together with signals in right posterior STS regions implicated in social perception, rather than engaging only classical higher-order semantic regions.
% Description encoding was more spatially selective and was concentrated in the dorsal-attention network, spanning superior parietal and intraparietal cortex (LH\_DorsAttn\_Post\_3, LH\_DorsAttn\_Post\_6, RH\_DorsAttn\_Post\_4, RH\_DorsAttn\_Post\_5), lateral occipitotemporal cortex (LH\_DorsAttn\_Post\_1), the bilateral frontal eye fields (LH\_DorsAttn\_FEF\_1, RH\_DorsAttn\_FEF\_1), and left ventral premotor cortex (LH\_DorsAttn\_PrCv\_1). Additional significant parcels lay in lateral and ventral occipital visual cortex (LH\_Vis\_2, LH\_Vis\_7, RH\_Vis\_2, RH\_Vis\_3, RH\_Vis\_7), in right posterior superior temporal sulcus extending to the temporo-parietal junction (RH\_Default\_Temp\_3, RH\_SalVentAttn\_TempOccPar\_1), in right anterior temporal cortex (RH\_Default\_Temp\_2), in right anterior insula (RH\_SalVentAttn\_FrOperIns\_1) and supplementary motor area (RH\_SalVentAttn\_Med\_2), in somatomotor cortex (LH\_SomMot\_1, RH\_SomMot\_5, RH\_SomMot\_7), and in the left parahippocampal gyrus (AAL). This pattern suggests that scene descriptions captured distributed visual and attentional information, together with signals in right posterior STS regions implicated in social perception, rather than engaging only classical higher-order semantic regions.

\subsubsection{Feature encoding follows distinct patterns of age-related differences}

\begin{figure}[!htbp]
    \centering
    \includegraphics[
    width=\textwidth]{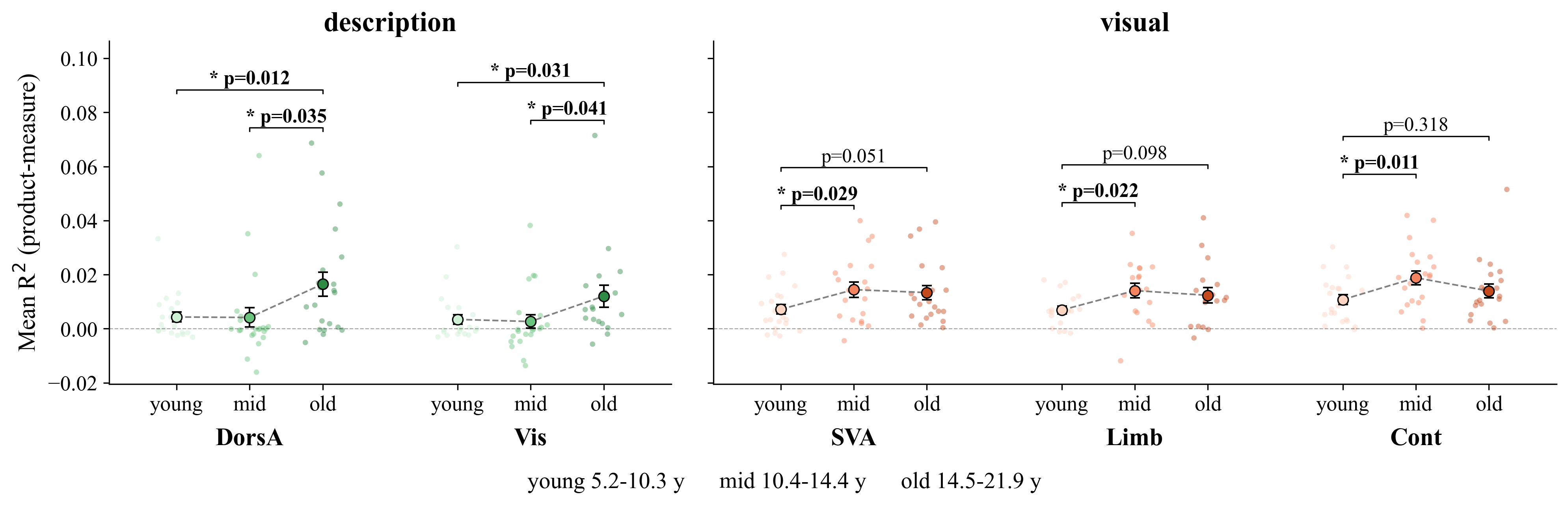}
    \caption{\textbf{Age-related changes in feature encoding, shown for the networks with the age effect.} Each point is one participant's product-measure $R^{2}$ averaged across the ROIs of a given network; large markers denote the tertile mean $\pm$ SEM, colour-shaded from light (young) to dark (old). Left: encoding of the \emph{description} feature (chunk length 20) in the dorsal attention (DorsA) and visual (Vis) networks, where the oldest tertile is markedly higher than both younger groups. Right: encoding of the low-level \emph{visual} feature (chunk length 16) in the salience/ventral-attention (SVA), limbic (Limb), and control (Cont) networks, where the middle and old tertiles exceed the young tertile. Brackets give two-tailed label-permutation $p$-values (10{,}000 permutations, uncorrected); an asterisk marks $p < 0.05$. Age tertiles: young 5.2--10.3~y, middle 10.4--14.4~y, old 14.5--21.9~y ($n = 22$ each).}

    \label{fig:encoding_age}
\end{figure}

To test whether encoding performance varied with age, we divided the 66 participants into three equal-sized age tertiles ($n = 22$ each): young (5.2--10.3~y), middle (10.4--14.4~y), and old (14.5--21.9~y). For each participant we averaged the product-measure $R^{2}$ across the ROIs within each of the seven cortical networks and an additional AAL ROI group, and compared tertiles using two-tailed label-permutation tests (10{,}000 permutations) at each feature\,$\times$\,network combination.
Encoding of the \emph{description} feature differed across age groups in the dorsal attention (DorsA) and visual (Vis) networks (Fig.~\ref{fig:encoding_age}). In both networks, the oldest tertile showed higher encoding than either younger tertile, and the two younger tertiles did not differ (DorsA: young $= 0.0043$, middle $= 0.0041$, old $= 0.0165$; old vs.\ young $p = 0.012$, old vs.\ middle $p = 0.035$. Vis: young $= 0.0034$, middle $= 0.0027$, old $= 0.0120$; old vs.\ young $p = 0.031$, old vs.\ middle $p = 0.041$; uncorrected, chunk length 20). The old-versus-young difference in DorsA had the same direction at all chunk lengths but did not reach significance at a chunk length of 16 TRs (Appendix Fig.~\ref{fig:encoding_des_age_app}).

The low-level \emph{visual} (AlexNet) feature showed no age-related difference within the visual network. In the salience/ventral-attention (SVA), control (Cont), and limbic (Limb) networks, however, visual encoding was higher in the middle than in the young tertile (SVA: young $= 0.0072$, middle $= 0.0144$, $p = 0.029$; Limb: young $= 0.0069$, middle $= 0.0141$, $p = 0.022$; Cont: young $= 0.0108$, middle $= 0.0188$, $p = 0.011$; uncorrected, chunk length 16), with no further increase in the old tertile.

\section{Discussion}
% brief summary results
We used naturalistic movie-watching fMRI data to examine whether multimodal large language models (MLLMs) can capture information relevant to brain activity during movie watching. 
First, we compared MLLM-generated socio-affective ratings (theory-of-mind and valence) against human ratings to assess how closely the two correspond. 
Second, we built a unified ROI-wise encoding model in which machine-generated features, including semantic scene descriptions, visual representations, and socio-affective ratings, were used to predict brain activity during movie watching. 
Overall, MLLM ratings tracked human ratings closely, and the machine-generated features explained meaningful neural responses whose spatial profile varied with age, indicating feature-specific developmental differences.

\subsection{Human–Machine Correspondence in Ratings and Neural Responses}
This study evaluated MLLMs as annotators of dynamic social and affective content by comparing their ToM and valence ratings with human judgments and with movie-evoked brain responses. The selected machine-generated time series showed partial correspondence with human ratings, suggesting some overlap in the social and affective information captured by the two sources. Our prior work~\cite{huang2026brain} compared models and context configurations in detail and showed that machine-generated ratings may offer a scalable alternative to human annotation in some settings, although variation across models, constructs, and movies limits their use as a general substitute. Here, we directly correlated each rating source with regional fMRI signals and examined the spatial overlap between the resulting maps, which asks whether agreement at the behavioral level is accompanied by shared regional associations with brain activity.

For ToM, the most consistent human–machine correspondence across both models was observed in bilateral lateral occipitotemporal cortex near MT+, rather than in early visual cortex. This location is compatible with the proposed third visual pathway, in which dynamic social information is processed from early visual cortex through motion-sensitive regions toward the STS ~\cite{pitcher2021evidence,mcmahon2023hierarchical}. Naturalistic movie studies further show that the STS is sensitive to a broad range of social cues, including faces, biological motion, goal-directed actions, emotion, and social interaction~\cite{lahnakoski2012naturalistic}. Beyond this shared component, the cross-movie overlap differed between models. LLaDA-V additionally overlapped with human ratings in a ventral posterior cingulate parcel of the default-mode network (negative correlation), whereas Qwen3-VL extended the positive overlap to an adjacent left lateral occipitotemporal parcel and additionally overlapped in medial occipital cortex (negative correlation).
% Beyond this shared lateral occipitotemporal component, the cross-movie overlap differed between models. LLaDA-V additionally overlapped with human ratings in a ventral posterior cingulate parcel of the default-mode network (negative correlation). Qwen3-VL instead extended the positive overlap to an adjacent left lateral occipitotemporal parcel assigned to the dorsal-attention network (LH_DorsAttn_Post_1) and additionally overlapped in medial occipital cortex (RH_Vis_6; negative correlation).

Human ToM ratings also correlated negatively with bilateral ventral posterior cingulate parcels of the default-mode network, regions consistently implicated in mentalizing~\cite{schurz2014fractionating, saxe2004understanding}.
These results show that these higher-order mentalizing regions can track moment-to-moment variation in ToM during naturalistic movie viewing in the opposite direction from the predominantly positive correlations observed in visual and attentional regions. This distinct correlation pattern may reflect a different role of default-mode regions in processing socially and cognitively complex movie content.
Together, these findings suggest that behavioral similarity between human and machine ToM ratings does not necessarily imply identical neural correspondence. Their shared judgments may be grounded in common perceptual cues available in the movie, while correspondence with higher-order systems appears to vary across models, with LLaDA-V matching human ratings in a mentalizing region and Qwen3-VL in a dorsal-attention region.
Human ratings may additionally reflect higher-order processes related to mental-state inference and contextual integration that are only partially captured by the MLLMs.

Valence showed a markedly more stimulus-dependent pattern than ToM. Within individual movies, human valence correlations were predominantly negative in TP and DM but almost entirely positive in PC, whereas ToM correlations were predominantly positive across all three movies; the negative pattern may reflect stronger responses to negatively valenced moments in the first two movies.
Human valence ratings identified only a single ROI that was consistently associated with BOLD activity across all three movies (left posterior dorsal attention network, positive correlation), and human--machine correspondence was similarly context dependent. In TP, both MLLMs reproduced the broad negative correlations that human ratings showed across default-mode regions; in DM this correspondence was only partial, and in PC the human ratings did not show this pattern. Across movies, the two models converged on a largely shared set of positively correlated visual and dorsal attention regions, with Qwen3-VL additionally involving salience/ventral attention and somatomotor regions; only Qwen3-VL shared the single human cross-movie-consistent ROI. This variability is consistent with valence being flexibly represented across distributed brain systems rather than localized to a fixed set of regions~\cite{lindquist2016brain}. Affective responses during naturalistic viewing can also vary substantially across individuals~\cite{chen2020intersubject}, and recent work further suggests that neural representations of moment-to-moment valence may be strongly context dependent and fail to generalize across different movies~\cite{ke2025dynamic}. Thus, although both MLLMs can recover human-aligned valence responses within particular movies, their correspondence with human ratings across stimulus contexts was minimal, suggesting that valence correspondence is more sensitive to both stimulus context and model-specific representations than ToM.

\subsection{Neural Encoding of MLLM-Derived Representations}

To assess the neural relevance of these machine-derived signals, we examined how these feature spaces contributed to the prediction of brain activity. A key feature of our encoding approach is that the description feature space consisted of natural-language scene descriptions generated by an MLLM, rather than predefined human annotations or low-level perceptual features extracted directly from the stimulus. Within a single banded-ridge model, the three feature spaces showed distinct, feature-dependent spatial profiles. Visual features extracted from AlexNet produced the strongest and most widespread encoding, extending across nearly all cortical parcels, whereas description features were concentrated in dorsal attention and visual regions, with additional effects in right lateral temporal cortex and the right temporo-parietal junction.
Previous voxel-wise encoding studies have described a gradient from lower-level visual representations in sensory cortex to higher-level semantic representations in association cortex~\cite{gucclu2015deep, huth2016natural}, and recent work has mapped AI-derived semantics at the ROI level~\cite{li2026mapping}. Our results at the coarser parcel scale show that machine-derived feature spaces can be distinguished by their spatial profiles, although they do not by themselves establish a perceptual-to-semantic hierarchy.
The description features here are derived from machine-\emph{generated} natural-language descriptions rather than from hand-labeled semantic categories or semantic representations derived from human language stimuli~\cite{huth2016natural}. That these descriptions explained variance mainly in dorsal attention and visual regions, with more limited effects in right lateral temporal cortex, suggests that they capture perceptual as well as semantic aspects of the movie rather than a purely higher-order semantic representation.

One consideration is that the MLLM-generated descriptions are largely grounded in the same movie content and may therefore retain substantial perceptual information. Because the visual and description feature spaces are partially correlated, some predictive variance is shared between the two bands and cannot be uniquely attributed to the semantic representation. Description encoding was also sensitive to feature dimensionality, becoming more widespread when the description space was increased from 64 to 128 dimensions (see Appendix~\ref{sec:description_dim} for details). In contrast, the visual encoding pattern was relatively stable, possibly because the visual representation already contained sufficient information at the dimensionality used here, whereas increasing the dimensionality of the description space allowed additional semantic variance to be captured. The stronger visual feature space may therefore account for part of the variance that is also present in the descriptions, potentially reducing the predictive contribution allocated to the description features. A similar issue may affect the social features: several socially relevant stimulus properties, such as the presence of faces and the number of people, were correlated with both the visual and description feature spaces. As a result, variance related to social content may already be explained by these richer representations, which could partly account for the small product-measure contribution of the social features in the encoding model.

\subsection{Developmental Differences in Feature Encoding}

Beyond spatial specificity, the machine-generated features were also developmentally sensitive. Encoding of the semantic description feature increased with age in dorsal-attention and visual networks, being strongest in the oldest participants, whereas the low-level visual feature showed no age-related change within early visual cortex but increased from the young age group into the middle one in the salience, control, and limbic networks.
In these latter networks, mean visual-feature encoding was numerically lower in the oldest than in the middle age group, suggesting a potentially nonmonotonic age relationship. A related pattern was reported by Di et al.~\cite{di2023individual}, who observed inverted-U-shaped age associations in the inter-individual consistency of movie-evoked brain responses in selected networks. Although their PCA-based consistency measure differs from our feature-based encoding measure, their findings provide context for the possibility that age-related differences in responses to naturalistic stimuli need not follow a monotonic increase.

These findings are broadly compatible with the sensorimotor-to-association axis of cortical development. Along this axis, primary sensory systems mature early while association systems continue to refine through adolescence~\cite{sydnor2021neurodevelopment, sydnor2023intrinsic}. The stability of the visual feature within early visual cortex is consistent with the early maturation of primary visual areas, where synaptogenesis and pruning reach adult levels years before association and prefrontal cortex~\cite{huttenlocher1997regional} and basic visual functions approach adult levels by early school age; higher-order, category-selective ventral-stream regions, in contrast, continue to develop through the teenage years~\cite{golarai2007differential, grill2008developmental}. That the visual feature's encoding nonetheless increased with age in the salience, control, and limbic networks indicates that it is not the early visual representations themselves that change, but rather the downstream integration of that visual information into attentional, executive, and affective systems that keeps maturing into adolescence. Prior developmental work has characterized this hierarchy through intrinsic or structural properties, or through the maturation of specific networks during movie watching~\cite{richardson2018development}; here the same axis emerges from the encoding strength of a machine-\emph{generated} semantic feature. That an automatically produced description of the ongoing narrative is sufficient to recover age-related change in higher-order cortex demonstrates that machine ratings are informative not only about where information is represented but also about how those representations develop.

\subsection{Considerations and Future Directions}

Several considerations should be kept in mind when interpreting these findings, and they also point to directions for future work. 
We evaluated two MLLMs and two socio-affective dimensions (ToM and valence), so future studies could extend this framework to a broader range of multimodal models (such as InternVL3~\cite{zhu2025internvl3} and GLM-4.5V~\cite{hong2025glm}) and social-cognitive constructs.
Our results also suggest that performance should be evaluated across multiple movie stimuli, particularly for valence, which showed greater variability across movies. In addition, future work could use richer and more multidimensional social features rather than a small set of scalar ratings, which may better capture the complexity of social information during naturalistic viewing.
Our dataset also involved relatively short movie clips, reflecting the practical trade-off between the number of participants and the amount of viewing time collected from each participant, especially in developmental studies where long scans can be difficult for young children.
Although our use of three movies and a relatively large sample provides broader coverage across participants and stimulus contexts, future work could systematically examine how scan duration and sample size affect the reliability and generalizability of machine-derived encoding results. 
% In addition, the human and machine ToM scales were anchored differently: the human scale started at $-10$ (``not at all''), whereas the machine prompt placed the absence of mental states at 0, reserved negative values for barriers to connection, and also emphasized emotional connection. In practice, however, negative machine ratings were rare (0--3.7\% of frames), so both measures varied mainly along the positive range of mental-state engagement; human--machine agreement thus reflects correspondence between closely related, though not identical, operationalizations of ToM.
Finally, an important future direction is to extend this framework to neurodevelopmental and clinical populations, such as individuals with autism spectrum disorder, where differences in social-information processing may provide a particularly informative test of whether MLLM-derived socio-affective features can characterize individual variation in naturalistic brain responses. Together, these considerations highlight several directions for extending MLLM-based automatic annotation in naturalistic neuroimaging.

Overall, our findings show that MLLM-generated socio-affective ratings capture substantial aspects of human judgments, while MLLM-derived representations also relate systematically to brain activity and reveal developmental differences in feature encoding. Together, these results suggest that MLLMs can serve as a scalable complement to human annotation for studying perceptual, semantic, and social processing during naturalistic viewing.

\section{Declarations}

\subsection{Availability of data and materials}

MRI and eye-tracking data were obtained from publicly available sources: the HBN dataset (\url{http://fcon_1000.projects.nitrc.org/indi/cmi_healthy_brain_network/}), and the OpenNeuro repository (\url{https://openneuro.org/datasets/ds000228}). The rating data are publicly available on the Open Science Framework (OSF) repository: (\url{https://osf.io/qky8p}). Our original code will be made publicly available upon acceptance of the paper.

\subsection{Competing interests}
The authors declare that they have no competing interests.

\subsection{Funding}
This work was supported by the DOE SEA-CROGS project (DE-SC0023191) and the AFOSR project (FA955024-1-0231) to Y.H., Z.Z., and M.X.; the U.S. National Institutes of Health (R15MH125332) to X.D.; and the New Jersey Governor’s Council for Medical Research and Treatment of Autism (CAUT25BRP005) to X.D.

\subsection{Authors' contributions}
Y.H.: Conceptualization, Methodology, Analysis, Visualization, and Writing. Z.Z.: Methodology and Analysis. R.R.: Data collection. B.B.: Review and editing. X.D.: Conceptualization, Data collection, Investigation, Supervision, and Review and editing. M.X.: Conceptualization, Methodology, Supervision, Project administration, Funding acquisition, and Review and editing.
% All authors read and approved the final manuscript.

\subsection{Acknowledgment}
We acknowledge George B. Hanna for assistance with behavioral data collection and the computing resources provided by the High Performance Computing (HPC) facility at NJIT.
% funding
% undergraduate 

\bibliography{ref}

\section{Appendix} 

\begin{figure}[htbp]
    \centering

    \centering
    \includegraphics[width=\textwidth]{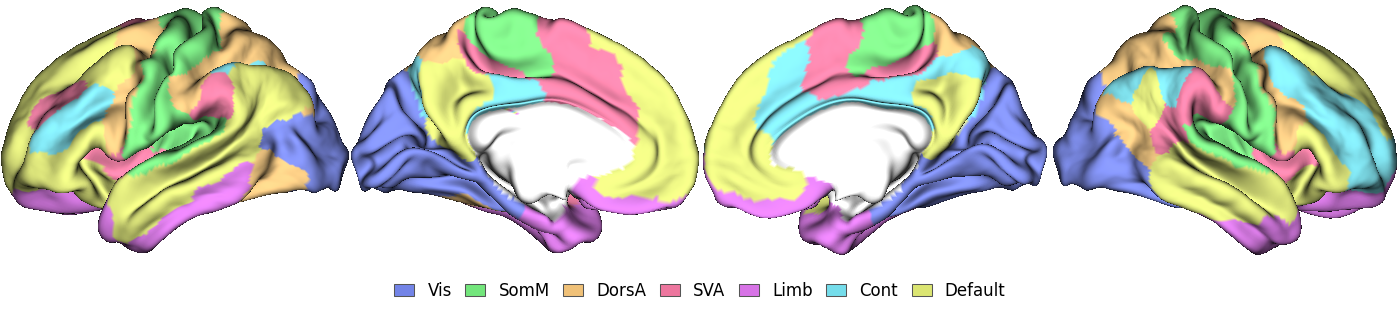}
    \caption{\textbf{Schaefer 100-parcel network atlas on the cortical surface.} The 100 cortical parcels of the Schaefer2018 atlas \citep{schaefer2018local} shown in their native fsLR (Conte69) surface space and colored by their 7-network assignment (Visual, Somatomotor, Dorsal Attention, Salience/Ventral Attention, Limbic, Control, and Default). Parcel labels were mapped to surface vertices with BrainSpace's \texttt{map\_to\_labels} and rendered on the Conte69 midthickness surface \citep{vos2020brainspace}. Lateral and medial views of the left and right hemispheres are displayed. The 14 subcortical AAL regions used in the analyses are not defined on the cortical surface and are therefore omitted from this rendering.}
    \label{fig:brain}
\end{figure}

\subsection{Encoding results}

\begin{figure}[!htbp]
    \centering
    \includegraphics[width=\textwidth]{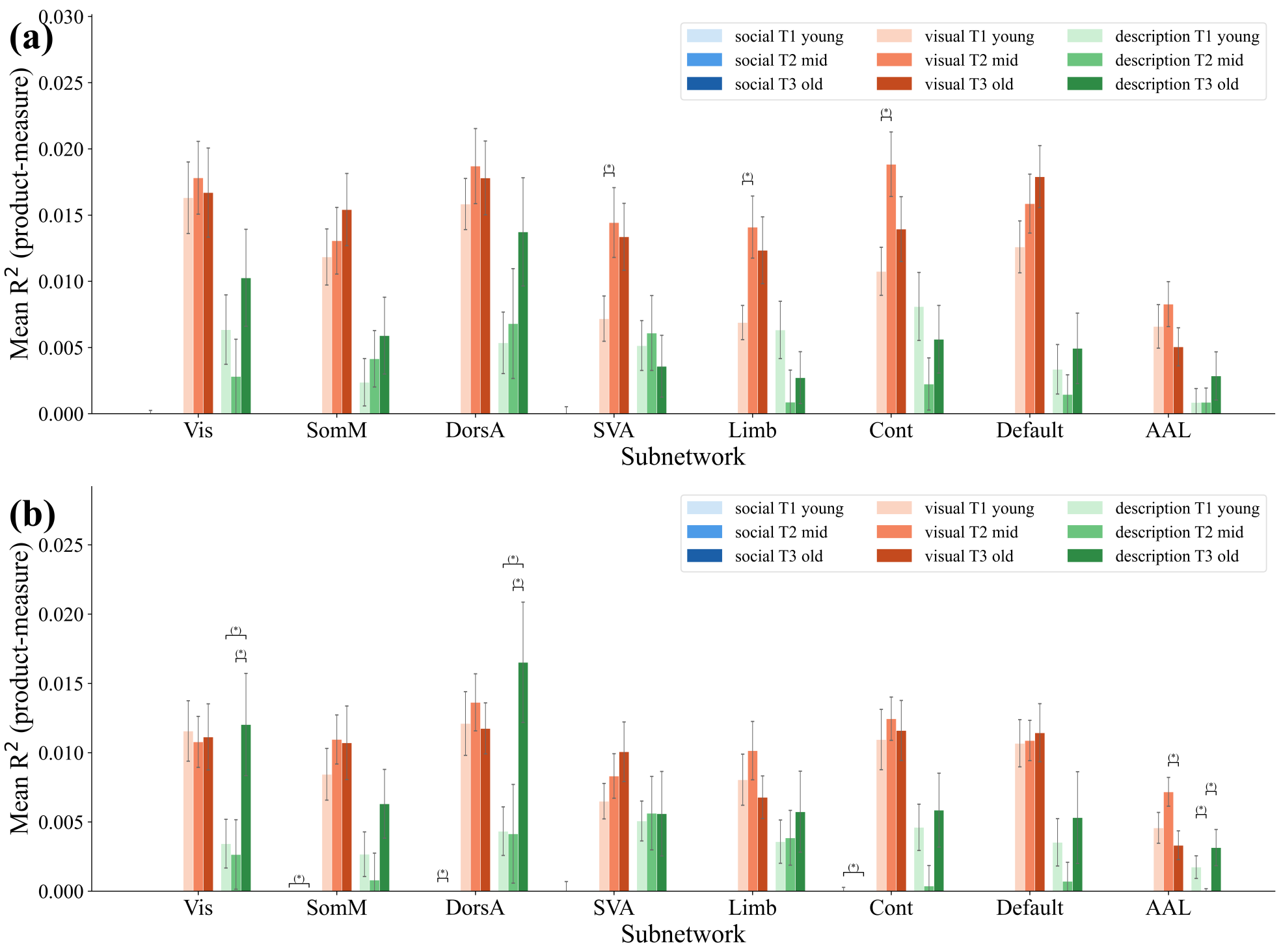}
    \caption{\textbf{Network-level encoding performance by feature space and age group.} Participants were divided into equal-sized age tertiles: younger (T1; 5.2–10.3 y, n = 22), middle (T2; 10.4–14.4 y, n = 22), and older (T3; 14.5–21.9 y, n = 22). Bars represent the mean product-measure R² for each feature space (social cognition, visual [layer 1], and language description) averaged across ROIs within each of seven cortical networks defined by the Schaefer 2018 100-parcel atlas, together with a subcortical (AAL) block. Error bars indicate ±1 standard error of the mean across participants. Asterisks (*) above brackets denote networks in which the oldest group (T3) differed significantly from the youngest (T1) or middle (T2) group, assessed by a two-tailed label-permutation test (10,000 permutations, uncorrected $p < 0.05$). Each group: n = 22. Visual dimension is 128 and description dimension is 64. (a) chunk length = 16, (b) chunk length = 20}
    \label{fig:encoding_bar_app}
\end{figure}

\subsubsection{Network-level age effects across all feature spaces}

Figure~\ref{fig:encoding_bar_app} shows the network-level encoding performance of all three feature spaces (social, visual, and description) across the three age tertiles, at two chunk lengths (cl16 in (a), cl20 in (b)). Several patterns are apparent. First, the social feature contributed negligibly in every network and every age group (bars near zero), consistent with its lack of reliable encoding at the single-subject level. Second, across most networks the visual (AlexNet) feature explained more variance than the description feature, indicating that low-level visual information dominated the overall encoding performance under naturalistic viewing. Third, within the visual (Vis) network the visual feature was essentially identical across the three tertiles, confirming that the low-level visual representation captured in early visual cortex is stable across this age range; the age-related increases in visual encoding instead emerged in higher-order networks (salience/ventral-attention, limbic, and control). By contrast, the description feature showed its age-related increase primarily in the dorsal-attention (and visual) networks. These patterns were consistent across the two chunk lengths, indicating that the developmental effects reported in the main text are not specific to a single cross-validation setting.

\begin{figure}[!htbp]
    \centering
    \includegraphics[width=\textwidth]{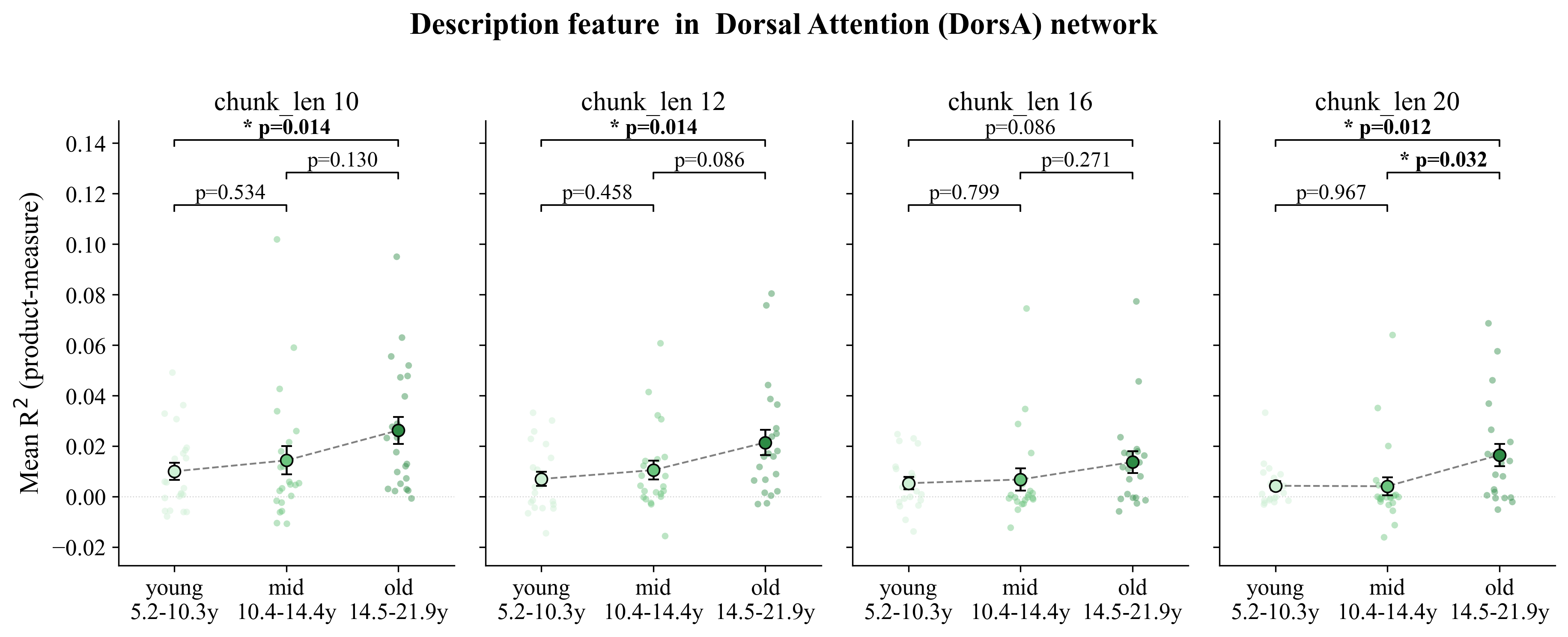}
    \caption{\textbf{Description--DorsA age effect across cross-validation chunk lengths.} Encoding performance for the description feature in the dorsal attention (DorsA) network, shown for four cross-validation temporal chunk lengths (chunk length = 10, 12, 16, 20 TRs; one panel each). In every panel participants are split into three equal-sized age tertiles (young, 5.2--10.3~y; mid, 10.4--14.4~y; old, 14.5--21.9~y; $n \approx 22$ each). Small semi-transparent dots show individual participants' network-mean product-measure $R^{2}$; large outlined markers show the group mean $\pm$ SEM, and the dashed line connects the group means. Brackets give all three pairwise comparisons (two-tailed label-permutation test, 10{,}000 permutations; $p$ uncorrected), with an asterisk marking $p < 0.05$. ($p$-values were estimated from 10,000 label permutations; small differences between figures ($\pm 0.003$) reflect Monte Carlo variability.) }
\label{fig:encoding_des_age_app}
\end{figure}

\subsubsection{Consistency of the description in DorsA network age effect across chunk lengths}

To verify that the age-related increase in description encoding within the dorsal-attention (DorsA) network was not an artifact of a particular cross-validation setting, we repeated the tertile analysis at four temporal chunk lengths used for movie-boundary-aware cross-validation (10, 12, 16, and 20 TRs). Figure~\ref{fig:encoding_des_age_app} shows the per-subject DorsA description $R^{2}$ for the three age tertiles at each chunk length. The pattern was consistent throughout: the oldest tertile showed the highest encoding at every chunk length (cl10: young $=0.010$, middle $=0.014$, old $=0.026$; cl12: $0.007 / 0.011 / 0.022$; cl16: $0.005 / 0.007 / 0.014$; cl20: $0.004 / 0.004 / 0.017$), and the old-versus-young difference was significant at three of the four chunk lengths (cl10 $p = 0.014$, cl12 $p = 0.014$, cl20 $p = 0.012$; two-tailed label-permutation, uncorrected) and marginal at cl16 ($p = 0.086$). Thus the developmental increase of description encoding in DorsA is robust to the choice of chunk length, with only its statistical strength varying across settings.

\begin{figure}[htbp]
    \centering
    \includegraphics[width=\textwidth, 
    trim={0 0 0 215},
    clip]{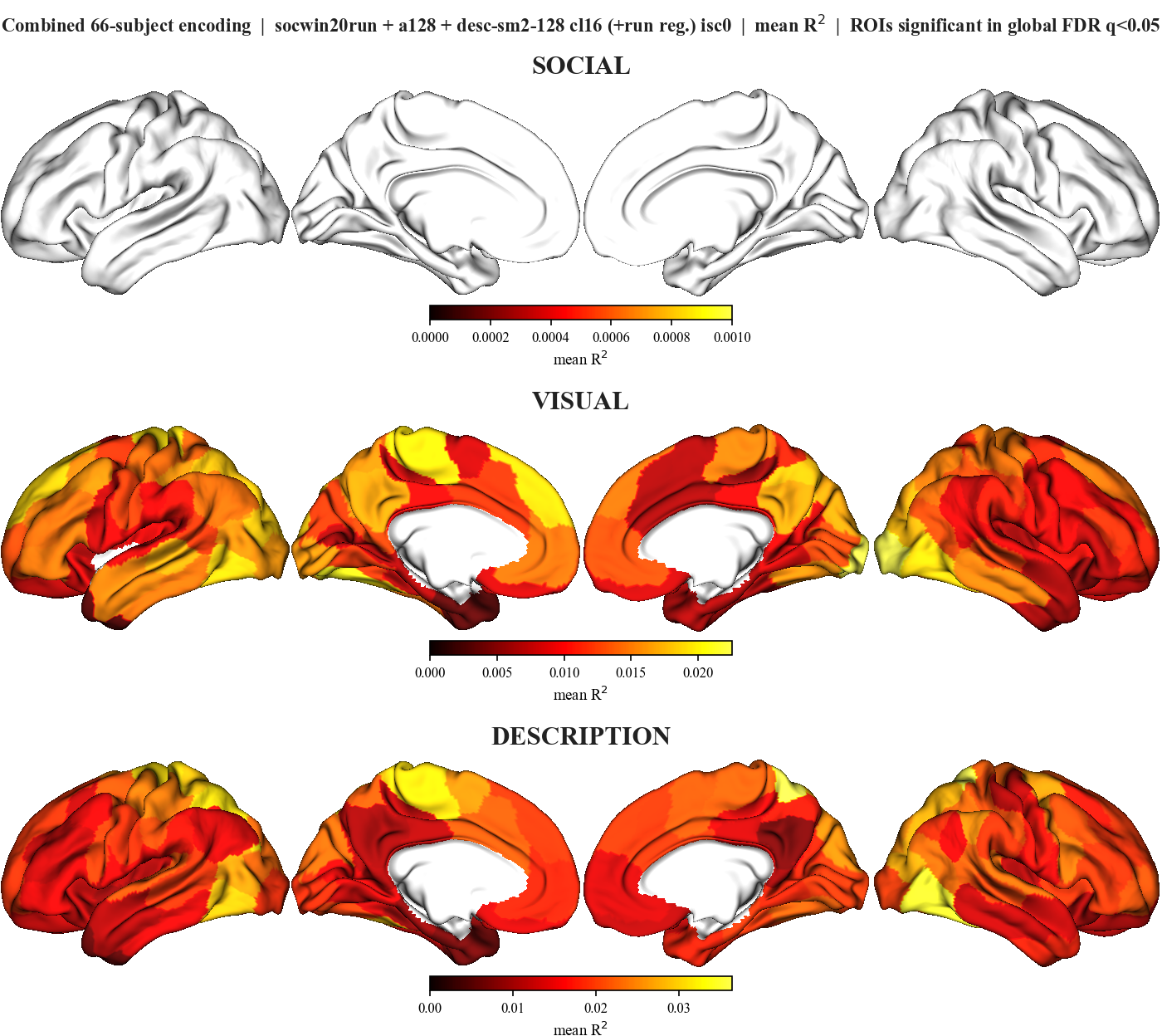}
    \caption{\textbf{Cortical surface maps of encoding performance for the visual (top) and description (bottom) features.} Encoding performance is quantified by the mean coefficient of determination ($R^{2}$), averaged across the full sample of 66 participants, and projected onto the Conte69 cortical surface with the \texttt{BrainSpace} library \citep{vos2020brainspace}. Parcel-wise $R^{2}$ values were computed on the Schaefer2018 100-parcel, 7-network atlas \citep{schaefer2018local} in its native fsLR (Conte69) surface space and assigned to surface vertices using BrainSpace's \texttt{map\_to\_labels} \citep{vos2020brainspace}.
    Colours indicate the parcel-wise mean $R^{2}$. Only parcels with significantly positive encoding after FDR correction (one-sided Wilcoxon signed-rank test against zero across the 66 participants, Benjamini--Hochberg $q < 0.05$) are coloured; non-significant parcels are shown in white. The \emph{social} feature is not shown because it did not survive FDR correction in any cortical parcel (0/100) and its mean $R^{2}$ was negative throughout. Subcortical and cerebellar regions are not displayed. Visual and description dimension are both 128, chunk length is 16.}
    \label{fig:brain_app}
\end{figure}

\subsubsection{Effect of description-feature dimensionality}
\label{sec:description_dim}

Because the semantic description feature is high-dimensional, we repeated the encoding analysis with the description space reduced to 128 dimensions (rather than 64 as in the main text) and found that the spatial pattern of encoding depended on this choice. Figure~\ref{fig:brain_app} shows the combined 66-subject maps for the 128-dimensional variant. The social feature again survived FDR correction in no cortical parcel, and the low-level visual (AlexNet) feature remained broadly encoded with a posterior maximum, as in the main analysis. The description feature, however, became far more widespread: it was significant across most cortical parcels, and its mean $R^{2}$ increased substantially. Notably, description encoding was now relatively high, and in places higher than visual encoding, in higher-order association networks, particularly the frontoparietal control network, in addition to the lateral-temporal and prefrontal regions seen at 64 dimensions.

Increasing the dimensionality of the description features from 64 to 128 was associated with higher encoding performance and a broader cortical distribution, including control and frontal regions. Because the description and visual feature spaces were partially correlated, this change may partly reflect differences in how shared predictive information was allocated between them. We report results using 64-dimensional description features in the main text and present the 128-dimensional results as a sensitivity analysis. These findings indicate that the strength and spatial extent of description encoding depend on feature dimensionality.

\subsection{Human Rating Prompts}
\definecolor{humangreen}{RGB}{46,139,87}
\outputfile{humangreen}{Theory of Mind Instructions}{prompts/human_tom.txt}

\outputfile{humangreen}{Emotional Valence Instructions}{prompts/human_valence.txt}

\subsection{Machine Rating Prompts}
\label{app:machine_prompt}
\outputfile{blue!60!black}{ToM rating prompt}{prompts/tom_prompt.txt}
\outputfile{forkorange}{Valence rating prompt}{prompts/valence_prompt.txt}
\outputfile{gray}{Scene description prompt}{prompts/description_prompt.txt}

\end{document}